\documentclass[aps,prl,twocolumn,superscriptaddress,nofootinbib]{revtex4-2}
\usepackage{tikz}
 \usetikzlibrary{arrows.meta,positioning,calc}
\usepackage[braket, qm]{qcircuit} 
\usepackage{graphicx}
\usepackage{amssymb}
 \usepackage{booktabs,array}
 \newcolumntype{L}[1]{>{\raggedright\arraybackslash}p{#1}}
\usepackage{indentfirst}
\usepackage{amsmath}
\usepackage{color,xcolor}
\usepackage{subfigure}
\usepackage{verbatim}
\usepackage[noend]{algpseudocode}
\usepackage{algorithm,algorithmicx}
\usepackage{amsthm}
\usepackage{lipsum}
\usepackage{enumitem}
\usepackage{amsmath}
\newtheorem{theorem}{Theorem}
\newtheorem{property}{Property}
\newtheorem{lemma}{Lemma}

\newtheorem{definition}{Definition}
\newtheorem*{remark}{Remark}
\usepackage{hyperref}
\hypersetup{
    colorlinks=true,
    linkcolor=blue,
    citecolor=blue,
    filecolor=blue,
    urlcolor=blue
}

\def\tr{\mathop{\mathrm{tr}}}

\begin{document}

\title{Quantum Advantage via Efficient Post-processing on Qudit Classical  Shadow Tomography}

\author{Yu Wang}
\email{wangyu@bimsa.cn}
\affiliation{Beijing Institute of Mathematical Sciences and Applications (BIMSA), Huairou District, Beijing 101408, P. R. China}

 \begin{abstract}
Computing inner products of the form \( \operatorname{tr}(AB) \), where \( A \) is a \( d \)-dimensional density matrix (with \( \operatorname{tr}(A) = 1 \), \( A \geq 0 \)) and \( B \) is a bounded-norm (BN) observable (Hermitian with \( \operatorname{tr}(B^2) \le O(\mathrm{poly}(\log d)) \) and \( \operatorname{tr}(B) \) known), is fundamental across quantum science and artificial intelligence. 
Classically, both computing and storing such inner products require $O(d^2)$ resources, which rapidly becomes prohibitive as $d$ grows exponentially. In this work, we introduce a quantum approach based on qudit classical shadow tomography, significantly reducing computational complexity from $O(d^2)$ down to $O(\mathrm{poly}(\log d))$ in typical cases and at least to $O(d~ \text{poly}(\log d))$ in the worst case. 
Specifically, for \(n\)-qubit systems (with $n$ being the number of qubit and \(d = 2^n\)), our method guarantees efficient estimation of \(\operatorname{tr}(\rho O)\) for any known stabilizer state \(\rho\) and arbitrary BN observable \(O\), using polynomial computational resources.
 Crucially, it ensures constant-time classical post-processing per measurement and supports qubit and qudit platforms. Moreover, classical storage complexity of $A$ reduces from $O(d^2)$ to $O(m \log d)$, where the sample complexity $m$ is typically exponentially smaller than $d^2$. Our results establish a practical and modular quantum subroutine, enabling scalable quantum advantages in tasks involving high-dimensional data analysis and processing.

\end{abstract}

\maketitle

 \textit{Introduction---}Computing \(\text{tr}(AB)\) for two known \(d\)-dimensional Hermitian operators is fundamental in quantum science and artificial intelligence. When \(A\) and \(B\) lack structure (e.g., sparsity or low rank), the classical computational and storage costs scale as \(O(d^2)\), which becomes prohibitive for exponentially large \(d\). 
In quantum settings, similar quantities arise: the expectation value \(\text{tr}(\rho O)\) predicts the outcome of measuring an observable \(O\) on quantum state \(\rho\).  
Such predictions are central to quantum information processing, quantum simulation, and quantum chemistry. However, when \(\rho\) is unknown and the dimension is exponential, estimating \(\text{tr}(\rho O)\) efficiently becomes significantly more challenging.

Fortunately, classical shadow tomography based on random Clifford measurements (Clifford-ST) provides a scalable approach for efficiently estimating \( \text{tr}(\rho O) \), when \( \rho \) is an unknown \( n \)-qubit state and \( O \) is a bounded-norm (BN) observable satisfying \( \text{tr}(O^2) \le O(\text{poly}(n)) \)~\cite{huang2020predicting}, and given that \( \text{tr}(O) \) is known. 
It yields a substantial quantum advantage in experimental learning tasks, exponentially reducing the sample complexity relative to classical methods~\cite{huang2022quantum}.

Quantum advantage plays a central role in quantum computing, with representative breakthroughs including the Deutsch–Jozsa algorithm~\cite{deutsch1992rapid}, Shor’s factoring algorithm~\cite{shor1994algorithms}, Grover's search algorithm~\cite{grover1996fast}, and the HHL algorithm~\cite{harrow2009quantum}. As large-scale matrix operations become increasingly common in quantum science and artificial intelligence, it is natural to ask whether shadow-based methods—originally designed for quantum state tomography—can be repurposed as scalable quantum subroutine algorithms for computing quantities such as \(\text{tr}(AB)\), delivering broader advantages in both sampling and computational complexity.

However, several challenges limit existing shadow estimation protocols. In Clifford-ST, each single-shot measurement yields an estimator \(\tilde{\rho}\), and post-processing involves computing \(\text{tr}(\tilde{\rho} O)\), which can be computationally expensive when \(\tilde{\rho}\) is dense and \(O\) lacks efficient stabilizer expression. In the worst case, the post-processing cost can become exponential. Recent work has extended Clifford-ST from qubit systems (\(d = 2^n\)) to qudits of odd prime power dimensions \cite{mao2025prl}. 
Meanwhile, protocols based on mutually unbiased bases (MUBs)—orthonormal bases with constant pairwise state overlap magnitude $1/\sqrt{d}$~\cite{wootters1989optimal}—have been proposed for BN observables~\cite{wang2024classical}. However, the existence of a complete set of $d + 1$ MUBs remains unresolved for general dimensions~\cite{horodecki2022five}, limiting their applicability in the most general settings. 
On near-term quantum devices, implementing random Clifford circuits requires \(O(n^2)\) decomposed gates. A workaround on optical quantum platforms is to project onto random stabilizer states, which are equivalent but require \(O(n^3 2^n)\) classical preprocessing~\cite{struchalin2021experimental}. 

These challenges highlight the need for alternative shadow estimation protocols that simultaneously reduce classical post-processing overhead, support general dimensionality, and remain preprocessing-efficient.

In this work, we introduce Dense Dual Bases Classical Shadow Tomography (DDB-ST), based on randomized projective measurements over $2d$ dense dual bases.  
As summarized in Fig.~\ref{fig:overview_final}, panel (a) shows the schematic pipeline common to shadow-based estimation, while panel (b) contrasts the scaling of different methods.  
Clifford-ST attains sampling efficiency but may incur exponential post-processing costs.  
By contrast, DDB-ST achieves constant-time post-processing per measurement, leading to an overall computational cost that scales linearly with the sample size, and it remains applicable to arbitrary dimension $d$. 
For BN observables, the worst-case sample complexity is $O(d\,\mathrm{poly}(\log d))$, while in typical cases it reduces to $O(\mathrm{poly}(\log d))$. 
In particular, for any known $n$-qubit stabilizer state $\rho$ and a BN observable $O$, the estimation requires $\mathrm{poly}(n)$ resources.

Beyond quantum state estimation, DDB-ST can serve as a modular subroutine 
for a broad range of quantum information and simulation tasks. 
Representative examples include fidelity estimation for device certification~\cite{flammia2011direct}, 
entanglement verification via witness observables~\cite{weilenmann2020entanglement,guhne2009entanglement}, 
and readout in variational quantum algorithms such as VQE~\cite{peruzzo2014variational} 
and QAOA~\cite{farhi2014quantum}. 
In quantum simulation, verification-type observables also appear in lattice gauge theory~\cite{kogut1979,kokail2019self}, 
where DDB-ST can efficiently estimate low-rank projectors such as ground-state fidelities or 
membership in low-energy subspaces. 
At the same time, each shadow snapshot produced by DDB-ST is extremely sparse, 
compressing storage from $O(d^2)$ for a full density matrix to $O(m \log d)$ with $m$ samples. 
This reduction helps mitigate post-processing overheads in memory and data movement, 
a challenge that is expected to become increasingly important as quantum experiments 
and data-intensive AI applications continue to scale. 
These features make DDB-ST a practical and versatile primitive 
for scalable quantum-enhanced data processing.

 \begin{figure}[t]
\centering
\textbf{(a) Protocol schematic}\vspace{1.2mm}

\resizebox{0.9\linewidth}{!}{%
\begin{tikzpicture}[
  >=Stealth, line width=0.9pt,
  every node/.style = {font=\footnotesize, align=center},
  block/.style = {draw, rounded corners=2pt, inner sep=2pt, outer sep=0pt,
                  minimum width=33mm, minimum height=8.8mm}
]
\node[block] (A) at (0,0) {unknown state $\rho$\\ BN observable $O$};
\node[block] (B) at (3.7,0) {random measurements\\\textit{Clifford / DDB}};
\node[block] (C) at (0,-1.48) {classical snapshots\\$\{\tilde{\rho}_1,\tilde{\rho}_2,\ldots\}$};
\node[block] (D) at (3.7,-1.48) {estimate property\\$\mathrm{tr}(\rho O)$};

\draw[->] (A) -- (B);
\draw[->] (C) -- (D);
\draw[->] (B.south) to[out=-102, in=78, looseness=0.8] (C.north);
\end{tikzpicture}%
}

\vspace{2.5mm}
\textbf{(b) Complexity comparison
}\vspace{1.2mm}

\scriptsize
\setlength{\tabcolsep}{3pt}  
\renewcommand{\arraystretch}{1.1}
\begin{tabular}{lccc}
\hline\hline
Method & Samples & Post-proc./sample & Dim. \\
\hline
Classical & N/A & $O(d^2)$ &  $d$ \\
Clifford-ST & $O(\mathrm{poly}(\log d))$ & up to $O(d^2)$ &
\begin{tabular}[c]{@{}c@{}}$d=2^n$\cite{huang2020predicting}\\ $p^n$ \cite{mao2025prl}\end{tabular} \\
DDB-ST & \begin{tabular}[c]{@{}c@{}}worst $O(d\,\mathrm{poly}(\log d))$\\ avg. $O(\mathrm{poly}(\log d))$\end{tabular}
& $O(1)$ &  $d$ \\
\hline\hline
\end{tabular}

\caption{Overview of the proposed framework.  
(a) Pipeline from inputs $\{(\rho,O)\}$ to the estimation of $\mathrm{tr}(\rho O)$. 
(b) Comparison of complexities. For dense $d{\times}d$ matrices, classical storage and direct evaluation cost $O(d^2)$. Throughout, the observable $O$ is assumed to be specified by its matrix elements in the computational basis.}
\label{fig:overview_final}
\end{figure}

\textit{Limitations of Clifford-ST---}Clifford-ST can be understood as an approximate quantum algorithm for estimating expectation values $\tr(\rho O)$ of BN observables. 
Its sample complexity scales with $\tr(O^2)$, and is therefore efficient whenever $\tr(O^2)\le \mathrm{poly}(n)$ for an $n$-qubit system. 
However, the overall runtime is governed by the cumulative cost of post-processing across all samples, which for Clifford-ST may vary drastically—from polynomial time in favorable cases to exponential time in the worst case.  

Efficient evaluation is possible only in special cases, such as when the shadow snapshots are sparse or the observable admits a decomposition into a polynomial number of Pauli or stabilizer terms. 
These cases, however, represent a measure-zero subset of BN observables. 
Consequently, despite its favorable sampling efficiency, Clifford-ST suffers from exponential post-processing overhead in general $d$-dimensional settings (see  Supplemental Material I for details).

 To avoid this overhead, we modify the set of shadow snapshots such that, given the representation of \(O\) in the computational basis, the computational cost of evaluating \(\tr(\tilde{\rho} O)\) remains constant for any \(d\)-dimensional observable.

 \textit{Snapshots with dense dual basis states---}
We define the following states:
\begin{equation}
\begin{cases}
    |\phi_{jk}^{\pm}\rangle = \frac{1 }{\sqrt{2}}(|j\rangle \pm|k\rangle), \\
    |\psi_{jk}^{\pm}\rangle = \frac{1 }{\sqrt{2}}(|j\rangle \pm i|k\rangle).
\end{cases}
\end{equation}

The new collection of snapshots comprises a total of \(2d^2 - d\) elements:
\begin{multline}\label{eq:2d2}
    \mathcal{S}_{\text{DDB}} = \{ P_t = |t\rangle\langle t|, ~~t = 0, \ldots, d-1; \\
    P_{jk}^{\pm} = |\phi_{jk}^{\pm}\rangle \langle \phi_{jk}^{\pm}|,  
    Q_{jk}^{\pm} = |\psi_{jk}^{\pm}\rangle \langle \psi_{jk}^{\pm}|; ~~0 \leq j < k \leq d-1 \}.
\end{multline}
These rank-1 projectors are informationally complete, as their linear span covers the entire space of \(d \times d\) Hermitian operators \(\mathbb{M}_d(\mathbb{C})\). This property makes such sets suitable for classical shadow tomography. 
 
An efficient algorithm with \(O(\log d)\) iterations has been developed to construct a unitary ensemble \(\mathcal{U}_{\text{DDB}} = \{U_j\}_{j=1}^{f(d)}\), containing the minimal number of elements required to span all rank-1 projectors in Eq.~\eqref{eq:2d2}~\cite{wang2024direct}. 
Here, \(f(d) = 2d\) when \(d\) is odd, and \(f(d) = 2d - 1\) when \(d\) is even. 
Each orthonormal basis \(\{U_j |k\rangle : k = 0, \dots, d-1\}\) is referred to as a Dense Dual Basis (DDB).

In \(n\)-qubit systems, each DDB circuit consists of a single Hadamard gate (optionally followed by a phase gate \(S\)) and a permutation gate, which can be realized with \(n\) generalized Toffoli gates, each decomposable into \(O(n^3)\) one- and two-qubit gates. Consequently, the total gate count is upper-bounded by \(O(n^4)\). Although this gate count is higher than that of Clifford circuits, random projections onto DDB states yield exponentially improved classical pre-processing efficiency compared to projections onto stabilizer states~\cite{struchalin2021experimental}, as detailed in the Supplementary Material II. 

Moreover, DDB-ST generalizes naturally to arbitrary finite dimensions. On \(n\)-qubit systems, the total number of DDB states is \(O(2^{2n})\), significantly fewer than the \(O(2^{n^2})\) stabilizer states. In the following, we investigate the explicit reconstruction channel of DDB-ST, along with its sample complexity and classical computational cost.

\begin{theorem}[Reconstruction channel for DDB-ST]
Let \(\rho = \sum_{j,k=0}^{d-1} \rho_{jk} |j\rangle\langle k|\), and define \(P_k = |k\rangle\langle k|\) as the projectors onto the computational basis.

\begin{itemize}
    \item For odd dimensions \(d\), each unitary \(U_j\) in the ensemble $\mathcal{U}_{\text{DDB}}$  is sampled uniformly with probability \(1/(2d)\).
    \item For even dimensions \(d\), the ensemble $\mathcal{U}_{\text{DDB}}$ is sampled, where the identity \(I\) (corresponding to the computational basis) is selected with probability \(2/(2d)\), and each remaining \(U_j\) with probability \(1/(2d)\).
\end{itemize}

The corresponding quantum channel \(\mathcal{M}\) takes the form:
\begin{equation}
    \mathcal{M}(\rho) = \frac{1}{2d} \left[\rho + \operatorname{tr}(\rho)\, I + (d-1) \sum_{k=0}^{d-1} \rho_{kk} P_k \right].
\end{equation}

Its inverse reconstruction channel \(\mathcal{M}^{-1}\) is given by:
\begin{equation}\label{eq:inverseddb}
    \mathcal{M}^{-1}(\rho) = 2d \left[\rho - \frac{d-1}{d} \sum_{k=0}^{d-1} \operatorname{tr}(\rho P_k)\, P_k \right] - \frac{\operatorname{tr}(\rho)}{d}\, I.
\end{equation}
\end{theorem}

All technical proofs and detailed derivations are deferred to the Supplemental Material. 

 \begin{property}
For any estimated state $\rho$ and $d$-dimensional Hermitian observable $O$ with known trace,  
the classical post-processing cost of a single-shot DDB-ST measurement is constant $O(1)$, provided $O$ is specified by its matrix elements in the computational basis. Each snapshot $\tilde{\rho}$ is extremely sparse (at most four nonzero entries up to the shift $-I/d$) and can be stored in $O(\log d)$ memory. 
\end{property} 

If the relevant matrix elements of \(O\) in the computational basis are not pre-stored but can be queried in polynomial time, 
then the per-sample post-processing cost of DDB-ST remains efficient. 
However, when \(O\) is specified via its Pauli decomposition, 
the advantage of constant-time post-processing disappears if \(O\) contains exponentially many nonzero Pauli terms.

\begin{theorem}[Performance Guarantee]
   In a \(d\)-dimensional Hilbert space, when using DDB-ST to predict the expectation value of any observable \(O\), the worst-case variance for each quantum state \(\sigma\) is bounded by:
\begin{equation}\label{worst}
 \lVert O_0 \rVert_\text{shadow}^2 = \max_{\sigma: \text{state}} \lVert O_0 \rVert_{\sigma}^2 \le 2d \, \operatorname{tr}(O_0^2),
\end{equation}
where \(O_0 = O - \frac{\operatorname{tr}(O)}{d} I\). If the unknown state is sampled randomly according to the Haar measure, the average variance is bounded by:
\begin{equation}\label{average}
    \lVert O_0 \rVert_{I/d}^2 \le 2 \, \operatorname{tr}(O_0^2).
\end{equation}
\end{theorem}

\begin{definition}[Approximately DDB-Average State]
   A state \(\rho\) is called approximately DDB-average if it satisfies the following condition: 
   \begin{equation} \label{eq:ddb-average}
       \left|\text{tr}(\rho |\phi\rangle\langle \phi|) - \frac{1}{d}\right| \leq \frac{O(\text{poly}(\log d))}{d},
   \end{equation}
   for all \(|\phi\rangle \in \mathcal{S}_{\text{DDB}}\).
\end{definition}

The maximally mixed state \(\rho = I/d\) satisfies \(\tr(\rho |\phi\rangle\langle \phi|) = 1/d\) for all \(|\phi\rangle \in \mathcal{S}_{\text{DDB}}\), and is thus exactly DDB-average.

\begin{figure}[htb]
\centering
\includegraphics[width=0.45\textwidth]{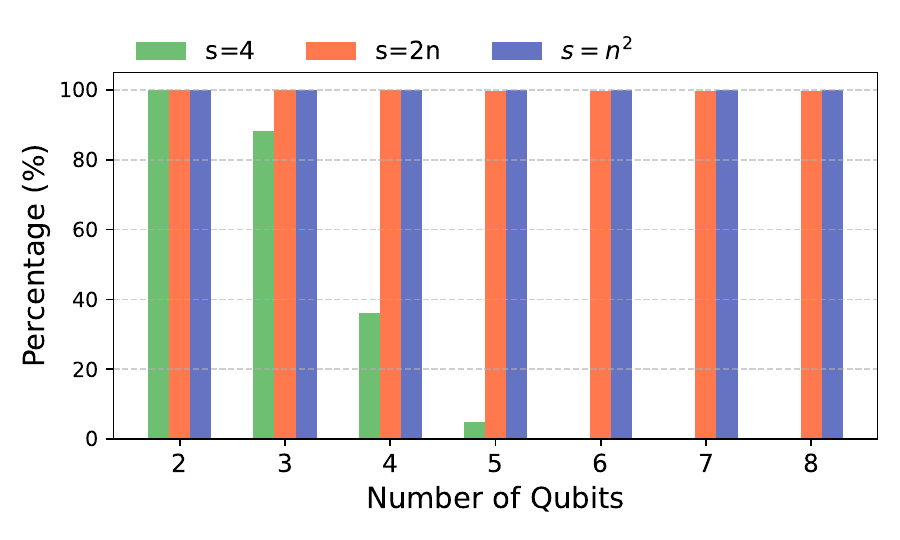}
\caption{Proportion of approximately DDB-average states.}
\label{percentage}
\end{figure}

We performed numerical simulations to estimate the prevalence of approximately DDB-average states (Fig.~\ref{percentage}). 
States are classified as approximately DDB-average if 
$\max_{|\phi\rangle \in \mathcal{S}_{\text{DDB}}} \left| \tr(\rho |\phi\rangle\langle\phi|) - \frac{1}{d} \right| \leq \frac{s}{2^n}$ 
for threshold parameter $s$. Using $10^3$ Haar-random pure states per dimension $d=2^2,\dots,2^8$, we found that the fraction of such states grows rapidly with $s$, approaching $100\%$ for $s=O(n^2)$. 
Interestingly, MUB-based classical shadow tomography has average variance 
$(1 + 1/2^n)\,\tr(O_0^2)$, about half that of DDB-ST. 
Consistently, in our numerical tests we found that thresholds scaling as 
$s=n$ for MUB and $s=2n$ for DDB were already sufficient to classify most 
Haar-random states as approximately average. 
Random mixed states from the Hilbert–Schmidt measure satisfy the DDB-average condition even more readily, with all samples passing the $s=5$ threshold for $n\leq8$.

\begin{lemma}
    For any BN-observable \( O \) and an approximately DDB-average state \( \rho \), the variance of DDB-ST satisfies:
    \(
    \lVert O_0 \rVert_{\rho}^2 \le O(\operatorname{poly}(\log d)) \cdot \operatorname{tr}(O_0^2).
    \)
    As a result, the sample and computational complexity for estimating \(\operatorname{tr}(\rho O)\) using DDB-ST is \( O(\operatorname{poly}(\log d)) \).
\end{lemma}

\textit{Complexity comparison---}  
For classical strategies, evaluating $\tr(AB)$ for general $d{\times}d$ inputs lacking special structure requires $O(d^2)$ resources in both time and storage. If, in addition, $A$ is generated by polynomial-depth quantum circuits containing sufficient non-Clifford (magic) gates, classical simulation of $A$ may incur overhead beyond $O(d^2)$, rendering the overall classical approach even less efficient.

For $n$-qubit systems, Clifford-ST achieves sample efficiency for BN observables—its sample complexity scales with $\operatorname{tr}(O^2)$, and is therefore polynomial whenever $\operatorname{tr}(O^2)\le \mathrm{poly}(n)$. 
However, the classical post-processing cost per measurement can vary drastically: while favorable instances admit polynomial overhead, in the worst case it can be as large as $O(4^n)$.

In contrast, DDB-ST guarantees constant $O(1)$ post-processing per measurement (Prop.~1), so the total computational complexity is linear in the sample complexity. 
Using the variance bounds in Eqs.~\eqref{worst}–\eqref{average}, the worst-case sample (computational) complexity is 
\[
N = O\!\left(\frac{\log \tfrac{1}{\sigma}}{\epsilon^2}\right)\, d\,\mathrm{poly}(\log d),
\]
while for typical states it reduces to 
\[
N = O\!\left(\frac{\log \tfrac{1}{\sigma}}{\epsilon^2}\right)\, \mathrm{poly}(\log d).
\]
Here $\epsilon$ is the target precision and $\sigma$ the failure probability. 

Consequently, DDB-ST yields exponential improvement whenever $\lVert O_0 \rVert_\rho^2 = O(\mathrm{poly}(\log d))$, and guarantees at least a near-quadratic speedup over the $O(d^2)$ classical baseline in the worst case (since $\lVert O_0 \rVert_{\text{shadow}}^2 \le 2d \operatorname{tr}(O_0^2)$). 
On average, polynomial computational complexity can be achieved when \(\rho\) is randomly chosen according to Haar measure for BN observable $O$. 
These complexity scalings are illustrated in the comparative schematic of Fig.~\ref{fig:overview_final}(b).

\textit{Exponential speedup examples}---While Eq.~\eqref{average} captures typical Haar behavior, many physically relevant settings involve structured states, where sharper complexity reductions arise. We therefore examine stabilizer states—central 
to quantum error correction and simulation \cite{gottesman1997stabilizer}—
motivating Property~2 and Theorem~3.

Any $n$-qubit stabilizer state can be expressed as a uniform superposition over 
an $r$-dimensional affine subspace $A \subset \mathbb{Z}_2^n$ with phases restricted to $\{1,-1,i,-i\}$,
\begin{equation}\label{eq:stabilizer}
   |\Psi\rangle = \frac{1}{\sqrt{2^r}} \sum_{k \in A} i^{\,q(k)} |k\rangle,
\end{equation}
where $0\le r \le n$, and $q:A\to\mathbb{Z}_4$ is a quadratic form 
\cite{dehaene2003clifford,nest2008classical}.

\begin{property}
For any stabilizer state $|\Psi\rangle$ expressed in Eq. (\ref{eq:stabilizer}), we have \begin{equation}
    \max_{|\phi\rangle \in \mathcal{S}_{\text{DDB}}} \text{tr}(|\Psi\rangle \langle \Psi |\phi\rangle\langle \phi|)  \le \frac{1}{2^r}. 
\end{equation} 
\end{property}

\begin{theorem}[Informal]
For any \(n\)-qubit stabilizer state \( |\Psi\rangle \) in Eq.~\eqref{eq:stabilizer} and BN observable \( O \), DDB-ST estimates \(\operatorname{tr}(|\Psi\rangle \langle \Psi| O)\) with additive error \( \epsilon + \sqrt{\operatorname{tr}(O^2)/2^r} \) using \( O(\mathrm{poly}(n)) \) samples and post-processing time. When \( O \) is off-diagonal, the additional 
term \(\sqrt{\operatorname{tr}(O^2)/2^r}\) disappears. 
\end{theorem}

This result does not contradict the Gottesman-Knill theorem~\cite{gottesman1997stabilizer}, which allows exact computation of \( \operatorname{tr}(\rho O) \) when both \( \rho \) and \( O \) are of stabilizer type. In contrast, our method extends efficient estimation to arbitrary BN observables \( O \), even when \( O \) lacks an efficient stabilizer decomposition. The cost is an additive error \( \epsilon + \sqrt{\operatorname{tr}(O^2)/2^r} \), where the second term decays exponentially with \( r \). For small $r$, the expectation value can also be computed directly using classical methods.

Beyond stabilizer states, DDB-ST provides exponential speedup for approximately uniform mixed states. Consider states of the form 
\(
\rho_A = \frac{I}{d} + \sum_{j \ne k} \rho_{jk} |j\rangle \langle k|\), with \(|\rho_{jk}| < \frac{\mathrm{poly}(\log d)}{d}.
\)
For any BN-observable \(O\), we have \(\left|\operatorname{tr}(\rho_A O) - \frac{\operatorname{tr}(O)}{d}\right| \le \frac{O(\text{poly}(\log d))}{d} \sum_{j\ne k}|O_{jk}|\le \text{poly}(\log d) \, \sqrt{\operatorname{tr}(O^2)}
\). 
 DDB-ST estimates this quantity with \(\epsilon\)-accuracy using \(O(\mathrm{poly}(\log d))\) computational resources. In contrast, \(\rho_A\) and \(O\) each involve \( O(d^2) \) variables, making it exponentially hard to perform the same estimation using purely classical computations. 

 A physically relevant example is the depolarizing channel $\mathcal{D}_p(\rho) = (1-p)\rho + p I/d$. For large $p$, the output state $\mathcal{D}_p(\rho)$ approaches the uniform mixture, making DDB-ST  efficient to evaluate expectation values $\operatorname{tr}[\mathcal{D}_p(\rho) O]$ 
in noisy quantum systems.

\textit{Applications beyond state learning---}
Beyond efficiently estimating the properties of quantum states, DDB-ST could serve as a general-purpose quantum subroutine for evaluating trace expressions of the form \(\operatorname{tr}(AB)\).

\begin{figure}
    \centering
    \includegraphics[width=1\linewidth]{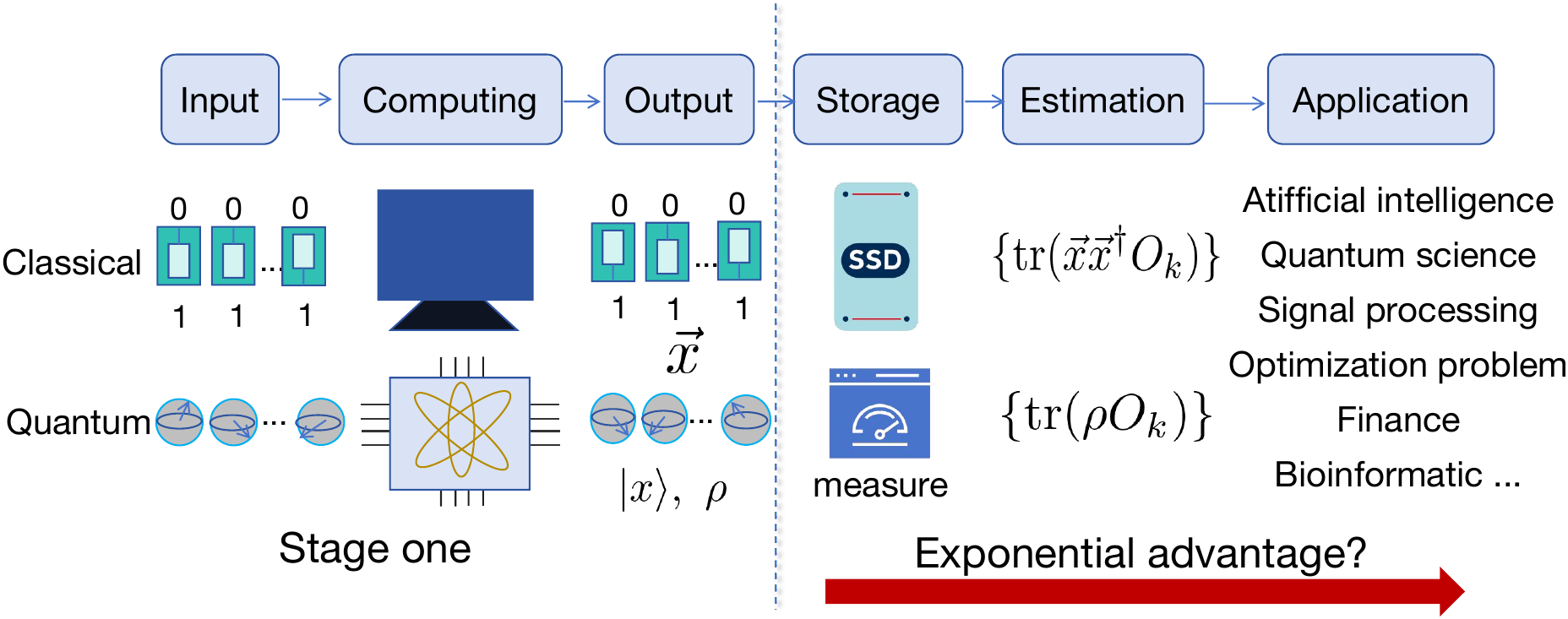}
    \caption{%
        Classical vs quantum estimation pipeline: enabling advantage with DDB-ST. 
        DDB-ST offers exponential or near-quadratic speedup in the estimation stage, bridging the gap between quantum state outputs and downstream applications in AI, optimization, and scientific computing \cite{rao1973linear,horn2012matrix}. 
        Although a quantum state \( \rho \) has limited lifetime, it could be compactly stored as polynomial-size classical data via well-designed random measurements by classical shadow tomography.
    }
    \label{fig:advantage}
\end{figure}

Such evaluations arise naturally in quantum algorithms where the output is a quantum state \( |x\rangle \), typically encoded in the amplitudes of a superposition and inaccessible via a single measurement. 
We assume that generating \( |x\rangle \) through quantum computation is not slower than classical counterparts, though many quantum algorithms seek exponential speedup at this stage. 
For instance, in the HHL algorithm~\cite{harrow2009quantum}, or its extensions in quantum machine learning~\cite{biamonte2017quantum}, the final state encodes the solution to \(A \vec{x} = \vec{b}\) as \( |x\rangle \), and estimating \(\operatorname{tr}(|x\rangle\langle x| O_k)\) for  observables \(\{O_k\}\) yields interpretable outputs for special application.  Efficiently performing this estimation is essential to realizing end-to-end quantum advantage, as shown in Fig. \ref{fig:advantage}.

\textit{Conclusion and discussion---}
In this work, we introduce DDB-ST, a shadow tomography framework using randomized projective measurements over $2d$ DDBs to estimate $\tr(AB)$ for $d$-dimensional density matrices $A$ and BN observables $B$. Our method achieves $O(d·\text{poly}(\log d))$ worst-case complexity and $O(\text{poly}(\log d))$ typical complexity, compared to $O(d^2)$ for direct classical 
trace computation. Unlike Clifford-ST which can suffer from exponential post-processing in the worst case, DDB-ST ensures $O(1)$ post-processing per measurement under the computational-basis representation of $O$, in  both qubit and qudit systems.
 Notably, our framework supports efficient estimation for arbitrary BN observables and $n$-qubit stabilizer states, further demonstrating its practical versatility.

There are several promising directions for future research. First, expanding the class of matrices \(A\) that allow for polynomial sampling complexity is crucial. This could involve biased sampling, imposing additional constraints on observables, or exploring sparse alternative snapshot sets. In \(n\)-qubit systems, DDB and MUB measurements represent two extremes of Clifford measurements. Nontrivial DDB and MUB states correspond to stabilizer states in Eq. (\ref{eq:stabilizer}) with \(r = 1\) and \(r = n\) respectively. Both DDB and MUB snapshots contain \(O(2^{2n})\) elements and exhibit similar variance properties. Including more snapshots with only polynomial nonzero amplitudes may reduce the worst-case sample complexity.

Second, while our work primarily focuses on the efficient computation of \(\tr(AB)\), extending these techniques to nonlinear properties, such as purity and entropy \cite{mcginley2023shadow}, could significantly broaden their range of applications. 

Third, many current quantum algorithms focus on achieving exponential speedups in Stage One (Fig.~\ref{fig:advantage}) using fixed quantum circuits. Incorporating measurements or randomized quantum circuits could unlock new potential for quantum advantage.  
In randomized measurement protocols, we design the unitary ensemble \( \mathcal{U}=\{U_j \}\), and nature replies with a measurement outcome \( |k\rangle \)—a stochastic echo from which we infer the properties of the quantum system.  
This interplay between controlled randomness and measurement-induced information reflects a deeper structure: even in non-determinism, nature leaves behind a traceable signature that is beneficial for learning.

Finally, improving the robustness and accuracy of the scheme in noisy environments is crucial \cite{chen2021robust,koh2022classical,wu2024error}. Since our primary goal is to estimate expectation values rather than fully reconstruct quantum states, the task would require less extensive error correction than typical quantum computations.

\textbf{Acknowledgements---}
This work received support from the National Natural Science Foundation of China through Grants No. 62001260 and No. 42330707, and from the Beijing Natural Science Foundation under Grant No. Z220002.

 \clearpage
\onecolumngrid
\appendix

\section{Classical shadow tomography framework with random Clifford measurements}

Classical shadow tomography, as introduced by Huang, Kueng, and Preskill~\cite{huang2020predicting}, provides a method for efficiently predicting properties of an unknown quantum state using randomized measurements. 

The process involves randomly sampling a unitary transformation \(U_k\) with probability $p_k$ from an informationally complete ensemble \(\mathcal{U} = \{U_k\}\), evolving the state \(\rho \to U_k \rho U_k^\dagger\), and performing projective measurements in the computational basis. 
It corresponds to a quantum process 
\begin{equation}
   \mathcal{M}(\rho)=\sum_{k,j} p_k \tr(U_k \rho U_k^{\dag} |j\rangle\langle j|) U_k^{\dag} |j\rangle\langle j| U_k.  
\end{equation} 
Each experiment yields a single estimator of the unknown state $\rho$:
\begin{equation}
 \tilde{\rho} = \mathcal{M}^{-1}\!\big(U_k^{\dagger} |j\rangle\langle j| U_k\big).
\end{equation} 
The estimation under observable \(O\) by one measurement is given by 
\begin{equation}\label{eq:post-processing}
\tr( \tilde{\rho}~ O).
\end{equation} 

Since the quantum state collapses upon measurement, multiple copies of the unknown state should be prepared. Given an exponentially large set of observables \(\{O_k\}_{k=1}^{L}\), to achieve accuracy \(\epsilon\) and confidence level \(1-\sigma\), we can ensure 
\begin{equation}
    P_r(|\langle o_k\rangle-\tr(\rho O_k)|<\epsilon)\ge 1-\sigma
\end{equation}
with the following sample complexity: 
\begin{equation}\label{eq:sampletime}
    N=O\left(\frac{\log \frac{L}{\sigma}}{\epsilon ^2}\right) \max_{1 \leq i \leq L} \left\| O_i - \frac{\text{tr}( O_i)}{d} I \right\|_{\mbox{shadow}}^2,
\end{equation}
where \(\|\cdot\|_{\mbox{shadow}}^2=\max_{\sigma: \text{ state }} \|\cdot\|_{\sigma}^2\) denotes the shadow norm associated with $\mathcal{U}$ and observables $\{O_k\}$. It represents the worst-case sample complexity, while \(\|\cdot\|_{\sigma}^2\) is related to the sample complexity for state \(\sigma\). 

If the ensemble \(\mathcal{U}\) consists of all \(n\)-qubit Clifford circuits, the inverse channel admits a closed-form:
\[
\mathcal{M}^{-1}(X) = (2^n + 1)X - \operatorname{tr}(X)\, I.
\]
The variance of Clifford-ST satisfies \(\|\cdot\|_{\text{shadow}} \le 3 \operatorname{tr}(O^2)\), ensuring that BN observables—those with \(\operatorname{tr}(O^2) \le \mathrm{poly}(n)\)—can be estimated using \(O(\mathrm{poly}(n))\) samples.

For each measurement, recording $U_k$ and outcome $j$, the post-processing computes 
\begin{equation}\label{eq:post-processing2}
\tr(\tilde{\rho}O)=(2^n + 1)\, \tr(U_k^{\dagger} |j\rangle\langle j| U_k O) -\, \tr(O),
\end{equation}
and averaging $N$ such measurements [cf. Eq.~\eqref{eq:sampletime}] yields the final estimate. 

The computational cost for Eq.~\eqref{eq:post-processing2} depends on the structure of \(\tilde{\rho}\) and \(O\). In Clifford-ST, the projected states, or snapshots, lie in the set
\[
\mathcal{S}_{\text{Clifford}} = \left\{ U_k^{\dagger} |j\rangle : U_k \in \mathcal{U}_{\text{Clifford}},\ j = 0,\dots,d-1 \right\},
\]
which contains \(O(2^{n^2})\) \(n\)-qubit stabilizer states. 
Due to the classical complexity of computing \(\tr(AB)\), evaluating \(\tr(\tilde{\rho} O)\) often requires \(O(2^{2n})\) operations when both matrices are dense.
 
Although the post-processing cost of Clifford-ST can be exponential in the worst case, there exist a few favorable scenarios where it becomes efficient.

\begin{itemize}
    \item \textbf{Sparsity of measurement states.}  
    If all collapsed states $\{U_k^{\dagger}|j\rangle\}$ are sparse with $O(\mathrm{poly}(n))$ nonzero amplitudes, then the evaluation
    \[
    \tr(\tilde{\rho} O) = (2^n+1)\,\tr(U_k^\dagger |j\rangle\langle j| U_k O) - \tr(O)
    \]
    can be computed in polynomial time for arbitrary $O$.  
    However, such sparse states form only a small fraction of the full Clifford snapshot set $\mathcal{S}_{\text{Clifford}}$.

    \item \textbf{Stabilizer- or Pauli-decomposable observables.}  Independent of the collapsed states,  
    if the observable $O$ admits a decomposition into a polynomial number of stabilizer projectors,
    \[
    O = \sum_{l=1}^{\mathrm{poly}(n)} c_l\,|\phi_l\rangle\langle \psi_l|,\quad
    |\phi_l\rangle,|\psi_l\rangle \in \mathcal{S}_{\text{Clifford}},
    \]
    or into a polynomial number of Pauli operators,
    \[
    O = \sum_{l=1}^{\mathrm{poly}(n)} \alpha_l\, P_l, \quad P_l \in \mathcal{P}_n,
    \]
    then the overlap $\tr(\tilde{\rho}O)$ can be computed efficiently using the Gottesman–Knill theorem, as stabilizer overlaps and Pauli expectation values on stabilizer states are efficiently computable. 
\end{itemize}

\noindent Unfortunately, both of these scenarios are highly restrictive: sparse measurement states constitute a negligible fraction of $\mathcal{S}_{\text{Clifford}}$, and observables with polynomial-size stabilizer or Pauli decompositions form a measure-zero subset of general BN observables.  

Thus, while Clifford-ST enjoys favorable sample complexity bounded by $\tr(O^2)$, its classical post-processing can still be exponential in the worst case, limiting its utility in high-dimensional settings for general BN observable.

\section{Efficient random projections in DDB-ST versus Clifford-ST}

\paragraph{Classical shadow tomography with random projective measurements or a single POVM.}
In the original formulation of classical shadow tomography~\cite{huang2020predicting}, one applies a random unitary $U$ drawn from an informationally complete ensemble $\mathcal{U} = \{U_k\}_{k=1}^L$ to the state $\rho$, followed by a computational basis measurement $\{|j\rangle\}_{j=0}^{d-1}$. 
Equivalently, a single shot of this procedure corresponds to a $d$-outcome projective measurement
\[
\{ U_k^\dagger |j\rangle\!\langle j| U_k \}_{j=0}^{d-1}.
\]
When this procedure is repeated many times with $U$ chosen uniformly at random, the overall measurement can be viewed as a single rank-1 informationally complete POVM with $Ld$ outcomes:
\[
\Bigl\{ \tfrac{1}{L}\, U_k^\dagger |j\rangle\!\langle j| U_k 
\;\Big|\; k=1,\dots,L;\ j=0,\dots,d-1 \Bigr\}.
\]

For example, the Pauli shadow scheme (randomly measuring in the $Z$, $X$, or $Y$ basis with equal probability) is equivalent to the six-outcome POVM
\[
\Bigl\{\tfrac{1}{3}|0\rangle\!\langle 0|, \ \tfrac{1}{3}|1\rangle\!\langle 1|, \ \tfrac{1}{3}|+\rangle\!\langle +|, \ \tfrac{1}{3}|-\rangle\!\langle -|, \ \tfrac{1}{3}|+i\rangle\!\langle +i|, \ \tfrac{1}{3}|-i\rangle\!\langle -i|\Bigr\}.
\] 
The POVM perspective has been emphasized in recent works~\cite{nguyen2022prl,innocenti2023prxq}, 
which show that shadow tomography can be more naturally and generally formulated in terms of generalized measurements. 
These studies highlight that the POVM viewpoint not only unifies different shadow protocols under a common framework, 
but also offers advantages in terms of generality, symmetry analysis, and optimization of measurement strategies. 

\paragraph{Clifford circuits versus DDB circuits.}
From the projective-measurement perspective, one should  randomly implement a unitary operation before performing the computational-basis measurement.  
The set of $n$-qubit Clifford circuits, generated by $\{CNOT, H, S\}$ gates, forms a finite group of cardinality
\[
|C_n| = 2^{\,n^2+2n} \prod_{j=1}^n (4^j - 1),
\]
which grows superexponentially with $n$.  
Direct uniform sampling from this set is highly nontrivial. Fortunately, efficient algorithms exist that can sample a random Clifford unitary in polynomial time~\cite{koenig2014efficient}.  
Moreover, each Clifford circuit admits a decomposition into $O(n^2)$ one- and two-qubit gates. 
However, on near-term intermediate-scale quantum (NISQ) devices, such quadratic gate counts quickly become impractical as $n$ grows, since gate errors and limited coherence times severely constrain the feasible circuit depth.

Clifford circuits are known to form an exact unitary 3-design in the qubit setting~\cite{zhu2017clifford3design}, a property that underlies their strong performance guarantees in classical shadow tomography. 
More recently, Schuster, Haferkamp, and Huang proved that random circuits on a variety of geometries---including one-dimensional layouts---can realize approximate unitary designs in depth $O(\log n)$ with optimal $n$-dependence~\cite{schuster2025science}. 
In particular, the construction yields $\varepsilon$-approximate 3-designs using one-dimensional log-depth Clifford circuits, and it is further showed that classical shadows with such log-depth Clifford circuits are as powerful as those with deep circuits, while requiring significantly reduced circuit depth.

Alternatively, a DDB circuit on an $n$-qubit system can be synthesized using a Hadamard (possibly followed by an $S$ gate) and a permutation.  
The permutation can be realized using at most $n$ generalized Toffoli gates, each of which decomposes into $O(n^3)$ one- and two-qubit gates.  
Consequently, the overall gate count for a DDB circuit is $O(n^4)$, which is asymptotically larger than that of Clifford circuits.  
 
\paragraph{Preprocessing complexity from the POVM perspective.}
In certain experimental platforms such as photonic systems, it is possible to directly implement rank-1 POVM projectors (e.g., via interferometric measurements) without executing a full random circuit, thus realizing classical shadow tomography in the POVM framework. 

For Clifford-based shadows, however, the associated POVM consists of all stabilizer projectors, with $O(2^{n^2})$ distinct elements. 
Under currently known constructions, uniformly generating a random $n$-qubit stabilizer projector is computationally expensive. 
A standard approach is to sample $n$ independent stabilizer generators $\{g_i\}_{i=1}^n$ and compute the corresponding projector as
\[
|\psi\rangle\langle \psi| = \frac{1}{2^n} \prod_{i=1}^n (I + g_i).
\]
The resulting stabilizer state vector has $2^n$ amplitudes, and a naive construction requires more than $O(2^{3n})$ computations~\cite[App.~A2]{struchalin2021experimental}. 

A more efficient alternative is to directly compute the stabilizer state in canonical form, specified by a triple $(R,t,q)$:
\begin{equation}\label{eq:stabilizer}
   |\Psi\rangle = \frac{1}{\sqrt{2^r}} \sum_{u \in \mathbb{Z}_2^r} i^{\,q(u)} \, |Ru+t\rangle ,
\end{equation}
where $R\in\mathbb{Z}_2^{n\times r}$ has full column rank, $t\in\mathbb{Z}_2^n$ is an offset vector, and $q:\mathbb{Z}_2^r\to\mathbb{Z}_4$ is a quadratic form. 
Equivalently,
\[
|\Psi\rangle = \frac{1}{\sqrt{2^r}} \sum_{k \in A} i^{\tilde q(k)} |k\rangle,
\]
with $A=\{Ru+t\}$ an $r$-dimensional affine subspace of $\mathbb{F}_2^n$. 
This representation reduces the preprocessing complexity to $O(2^n n^3)$~\cite[App.~Thm.~4]{struchalin2021experimental}, but the scaling remains exponential in~$n$.

\subparagraph{Preprocessing cost of sampling DDB projectors.} 

By contrast, the total number of different DDB states in a \(d\)-dimensional system is \(2d^2 - d\), and sampling from these states is straightforward. 
In the designed sampling strategy, each computational basis state \(|j\rangle\) is chosen with probability \(2/(2d^2)=1/d^2\), while every other nontrivial DDB state is chosen with probability \(1/(2d^2)\). We can sample a random projection onto a DDB state in the following way.

First, we select a computational basis  probability \(1/d\). Each state \(|j\rangle\) (where \(j = 0, \dots, d-1\)) is then sampled with probability $1/d$. Thus  \(|j\rangle\)  is selected with probability \(1/d^2\). 

Next, with probability \(1 - 1/d\), we select a non-computational basis. In this case, two distinct integers \(m, n \in \{0, \dots, d-1\}\) are selected such that \(0 \le m < n \le d-1\). Then, one of the following four superpositions is chosen randomly with equal probability (\(1/4\) each):
\[
\frac{1}{\sqrt{2}}\left(|m\rangle + |n\rangle\right), \quad \frac{1}{\sqrt{2}}\left(|m\rangle - |n\rangle\right), \quad \frac{1}{\sqrt{2}}\left(|m\rangle + i|n\rangle\right), \quad \frac{1}{\sqrt{2}}\left(|m\rangle - i|n\rangle\right).
\]

Thus, the probability of selecting any nontrivial DDB state is:
\[
\frac{1 - 1/d}{\binom{d}{2}} \times \frac{1}{4} = \frac{1}{2d^2}.
\]

In this process, each DDB state is selected according to the designed probability distribution. The computational resources required are only \(O(1)\). In contrast to Clifford-ST, where the projection state is determined through computationally intensive methods, the preprocessing overhead in DDB-ST is exponentially smaller. This is primarily because only two nonzero amplitudes need to be determined, while all other components are zero, eliminating the need for further computation.

\section{Proof of Theorem 1: calculation of the reconstruction channel}

We may express $\rho$ with the following form
\begin{equation}
    \rho=\sum_{j,k=0}^{d-1}\rho_{jk}|j\rangle\langle k|.
\end{equation}

The DDB unitary ensemble is $\{U_j\}_{j=1}^{f(d)}$. 
For even \(d\), \(f(d) = 2d - 1\). The first DDB corresponds to the computational basis, while the remaining bases are constructed in dual pairs, denoted as \(\{|\phi_{jk}^{\pm}\rangle : (j, k) \in \mathbb{T}\}\) and \(\{|\psi_{jk}^{\pm}\rangle : (j, k) \in \mathbb{T}\}\), where \(\mathbb{T}\) is a partition of \(\{0, 1, \dots, d - 1\}\) into distinct pairs with no repeated elements. For example, when \(d = 4\), one possible partition is \(\mathbb{T}_1 = \{(0,1), (2,3)\}\). A minimum of \(d - 1= \frac{C_d^2}{d/2}\) such partitions is needed to cover all pairs \((j, k)\) where \(0 \le j < k \le d - 1\).

For odd \(d\), \(f(d) = 2d\). In this case, the DDBs are similarly paired as \(\{|\phi_{jk}^{\pm}\rangle, |l_{\mathbb{T}}\rangle : (j, k) \in \mathbb{T}\}\) and \(\{|\psi_{jk}^{\pm}\rangle, |l_{\mathbb{T}}\rangle : (j, k) \in \mathbb{T}\}\), where \(l_{\mathbb{T}}\) represents the single element in \(\{0, 1, \dots, d - 1\}\) not included in \(\mathbb{T}\). For example, with \(d = 5\) and \(\mathbb{T}_1 = \{(0,1), (2,3)\}\), the value of \(l_{\mathbb{T}}\) is 4. A minimum of \(d= \frac{C_d^2}{(d-1)/2}\) such partitions is also sufficient. 

When $d$ is even, there are $2d-1$ DDBs, with the computational basis states $\{|0\rangle,|1\rangle,\cdots,|d-1\rangle\}$ sampled  twice. When $d$ is odd, there are $2d$ DDBs but 
the computational states $\{|t\rangle:t=0,\cdots,d-1\}$ appear twice.  Thus for general dimension $d$, the quantum channel for randomly sampling DDBs is given by: 
 
 \begin{align}
   2d \times \mathcal{M}(\rho)&= 2\sum_{k=0}^{d-1} \text{tr}(\rho P_k)  P_k +\sum_{0\le j<k\le d-1} 
   \left[ \text{tr}(\rho P_{jk}^{\pm}) \cdot P_{jk}^{\pm} + \text{tr}(\rho Q_{jk}^{\pm}) \cdot Q_{jk}^{\pm} \right] \label{4} \\ 
   &=2 \sum_{k=0}^{d-1}\rho_{kk}|k\rangle\langle k|+  \sum_{0\le j<k\le d-1} [ (\rho_{jj}+\rho_{kk})(|j\rangle\langle j|+|k\rangle\langle k|)  + \rho_{jk} |j\rangle\langle k|+  \rho_{kj} |k\rangle\langle j|  ] \label{5}\\
   &= \sum_{k=0}^{d-1}\rho_{kk}|k\rangle\langle k|+ \rho  +\sum_{0\le j<k\le d-1}  (\rho_{jj}+\rho_{kk})(|j\rangle\langle j|+|k\rangle\langle k|)\\
    &= \sum_{k=0}^{d-1}\rho_{kk}|k\rangle\langle k|+ \rho  +\frac{1}{2}[\sum_{j=0} ^{d-1} \sum_{k=0} ^{d-1} (\rho_{jj}+\rho_{kk})(|j\rangle\langle j|+|k\rangle\langle k|)-\sum_{k=0}^{d-1}(\rho_{kk}+\rho_{kk})(|k\rangle\langle k|+|k\rangle\langle k|)]\\
     &= \sum_{k=0}^{d-1}\rho_{kk}|k\rangle\langle k|+ \rho  +\frac{1}{2}[ \sum_{j=0}^{d-1}(d \rho_{jj}|j\rangle\langle j|+\rho_{jj} I+\tr(\rho)|j\rangle\langle j|+\sum_{k=0}^{d-1}\rho_{kk}|k\rangle\langle k|)-4\sum_{k=0}^{d-1}\rho_{kk}|k\rangle\langle k|]\\
     &= \sum_{k=0}^{d-1}\rho_{kk}|k\rangle\langle k|+ \rho  +\frac{1}{2}(2d \sum_{k=0}^{d-1}\rho_{kk}|k\rangle\langle k|+2\mbox{tr}(\rho)I-4\sum_{k=0}^{d-1}\rho_{kk}|k\rangle\langle k|)\\
   &=  \rho+\mbox{tr}(\rho)I+(d-1)\sum_{k=0}^{d-1}\rho_{kk}|k\rangle\langle k|. 
  \end{align}

From Eq.(\ref{4}) to Eq.(\ref{5}), we use 
\begin{equation}\label{equ:four}
\begin{cases}
    \text{tr}(\rho P_{jk}^{+}) = (\rho_{jj}+\rho_{kk}+\rho_{jk}+\rho_{kj})/2, \\
    \text{tr}(\rho P_{jk}^{-}) = (\rho_{jj}+\rho_{kk}-\rho_{jk}-\rho_{kj})/{2}, \\
    \text{tr}(\rho Q_{jk}^{+}) = (\rho_{jj}+\rho_{kk}+i\rho_{jk}-i\rho_{kj})/{2}, \\
    \text{tr}(\rho Q_{jk}^{-}) = (\rho_{jj}+\rho_{kk}-i\rho_{jk}+i\rho_{kj})/{2},
\end{cases}
\end{equation}
and 
\begin{equation}
\begin{cases}
    P_k = |k\rangle\langle k|,\\
     P_{jk}^{+} = (|j\rangle\langle j|+|k\rangle\langle k|+|j\rangle\langle k|+|k\rangle\langle j|)/2, \\
    P_{jk}^{-} = (|j\rangle\langle j|+|k\rangle\langle k|-|j\rangle\langle k|-|k\rangle\langle j|)/2, \\
     Q_{jk}^{+} = (|j\rangle\langle j|+|k\rangle\langle k|-i|j\rangle\langle k|+i|k\rangle\langle j|)/2, \\
     Q_{jk}^{-} = (|j\rangle\langle j|+|k\rangle\langle k|+i|j\rangle\langle k|-i|k\rangle\langle j|)/2. 
\end{cases}
\end{equation}
Denote the following symbols.  
\begin{equation}
\begin{cases}
     a_{jk} = \rho_{jj}+\rho_{kk}, \\
    b_{jk}^{\pm} = \rho_{jk}\pm\rho_{kj}, \\
    A_{jk} = |j\rangle\langle j|+|k\rangle\langle k| \\
    B_{jk}^{\pm} = |j\rangle\langle k|\pm|k\rangle\langle j|.\\
\end{cases}
\end{equation}
Thus we have 
\begin{equation}
\begin{cases}
     \tr(\rho P^{+}_{jk})P^{+}_{jk} =\frac{1}{4} (a_{jk}+b_{jk}^{+})(A_{jk}+B_{jk}^{+})\\
     \tr(\rho P^{-}_{jk})P^{-}_{jk} =\frac{1}{4} (a_{jk}-b_{jk}^{+})(A_{jk}-B_{jk}^{+})\\
     \tr(\rho Q^{+}_{jk})Q^{+}_{jk} =\frac{1}{4} (a_{jk}+i b_{jk}^{-})(A_{jk}-i B_{jk}^{-})\\
     \tr(\rho Q^{-}_{jk})Q^{-}_{jk} =\frac{1}{4} (a_{jk}-i b_{jk}^{-})(A_{jk}+i B_{jk}^{-}).\\
\end{cases}
\end{equation}

It is easy to verify that the summation above is equal to $a_{jk}A_{jk}+(b_{jk}^{+}B^{+}_{jk}+b_{jk}^{-}B^{-}_{jk})/2=(\rho_{jj}+\rho_{kk})(|j\rangle\langle j|+|k\rangle\langle k|)  + \rho_{jk} |j\rangle\langle k|+  \rho_{kj} |k\rangle\langle j| $.

Thus the quantum channel is given by  
\begin{equation}
    \mathcal{M}(\rho)=\frac{1}{2d}[ \rho+\mbox{tr}(\rho)I+(d-1)\sum_{k=0}^{d-1}\rho_{kk}|k\rangle\langle k|].
\end{equation}

As a comparison, when we use the uniform sampling of Clifford measurements or MUB measurements, the quantum channel is given by $\mathcal{M}(\rho)=\frac{1}{d+1}(\rho+\tr(\rho)I)$, where $d=2^n$. 

For the channel corresponding to uniform sampling from Cliffords and MUBs, the mapping of the matrix elements of a density matrix \(\rho\) is as follows:
\begin{itemize}
    \item The off-diagonal elements \(\rho_{jk}\) (where \(j \ne k\)) are mapped to \(\rho_{jk}/(d+1)\). 
   \item The diagonal elements \(\rho_{jj}\) are mapped to \(\frac{\rho_{jj} + \mathrm{tr}(\rho)}{d+1}\).
\end{itemize}

In contrast, for the channel corresponding to uniform sampling from DDBs, the mapping is: 
\begin{itemize}
    \item The off-diagonal elements \(\rho_{jk}\) (where \(j \ne k\)) are mapped to \(\rho_{jk}/(2d)\).
    \item  The diagonal elements \(\rho_{jj}\) are mapped to \(\frac{d \times\rho_{jj} + \mathrm{tr}(\rho)}{2d}\).
\end{itemize}
 
The inverse reconstruction channel of uniform sampling DDBs is given by 
 \begin{align}\label{inverse}
     \mathcal{M}^{-1}(\rho)
     &=2d\Bigg[\rho-\frac{d-1}{d}\sum_{k=0}^{d-1}\mbox{tr}(\rho P_k)P_k  \Bigg]-\frac{\mbox{tr}(\rho)}{d}I.
 \end{align}
It also includes the linear combination of \(\rho\) and \(I\), with additional corrections to the diagonal terms.


\section{Proof of Proposition 1: constant-time post-processing and storage analysis in a single measurement for DDB-ST} 

One of the most significant advantages of DDB-ST is that the classical post-processing per single-shot measurement has constant complexity $O(1)$. This result assumes that the observable $O$ is specified by its matrix elements in the computational basis, $O = \sum_{m,n=0}^{d-1} O_{mn} |m\rangle\langle n|$. Under this representation, each post-processing step involves accessing at most four matrix elements $O_{mn}$ and requires at most four arithmetic operations when $\operatorname{tr}(O)/d$ is deferred to the final averaging step.

\textit{Proof.}
We evaluate the post-processing formula
\[
\operatorname{tr} \left[ \mathcal{M}^{-1}(U_k^{\dagger} |j\rangle\langle j| U_k) \cdot O \right],
\]
where \(\mathcal{M}^{-1}\) is the inverse channel defined by Eq. (\ref{inverse}) 
with \(P_k = |k\rangle\langle k|\) being projectors onto the computational basis. 
Let \(\rho = U_k^{\dagger} |j\rangle\langle j| U_k = |\psi\rangle\langle\psi|\), where \(|\psi\rangle \in \mathcal{S}_{\text{DDB}}\) is the snapshot vector. 
We express the observable as \( O = \sum_{m,n=0}^{d-1} O_{mn} |m\rangle \langle n| \). 

There are three cases:

\textbf{(1) Computational basis state:}
\[
|\psi\rangle = |t\rangle \quad \Rightarrow \quad \rho = |t\rangle\langle t|.
\]
Then:
\[
\operatorname{tr}(\rho P_k) = \delta_{tk}, \quad \operatorname{tr}(\rho) = 1,
\]
so
\[
\sum_{k=0}^{d-1} \operatorname{tr}(\rho P_k) P_k = P_t = |t\rangle\langle t|,
\]
and thus:
\begin{equation}\label{constant1}
    \mathcal{M}^{-1}(\rho) = 2d \left[ |t\rangle\langle t| - \frac{d-1}{d} |t\rangle\langle t| \right] - \frac{1}{d} I = 2 |t\rangle\langle t| - \frac{1}{d} I.
\end{equation}
Hence:
\[
\operatorname{tr}[\mathcal{M}^{-1}(\rho) \cdot O] = 2 O_{tt} - \frac{\operatorname{tr}(O)}{d}.
\]

\textbf{(2) Real superposition state:}
\[
|\psi\rangle = \frac{1}{\sqrt{2}} (|m\rangle + |n\rangle),\quad m \ne n.
\]
Then:
\[
\rho = |\psi\rangle\langle\psi| = \frac{1}{2} \left( |m\rangle\langle m| + |n\rangle\langle n| + |m\rangle\langle n| + |n\rangle\langle m| \right),
\]
and
\[
\operatorname{tr}(\rho P_k) = \begin{cases}
\frac{1}{2}, & k = m \text{ or } n, \\
0, & \text{otherwise}.
\end{cases}
\]
So:
\[
\sum_k \operatorname{tr}(\rho P_k) P_k = \frac{1}{2} (P_m + P_n),
\]
and
\begin{equation}\label{constant2}
    \mathcal{M}^{-1}(\rho) = 2d \left[ \rho - \frac{d-1}{2d} (P_m + P_n) \right] - \frac{1}{d} I =P_m+P_n+d(|m\rangle\langle n| + |n\rangle\langle m|)- \frac{1}{d} I. 
\end{equation}
 
With the expression of $O$, we have   
\[
\operatorname{tr}[\mathcal{M}^{-1}(\rho) \cdot O] = O_{mm} + O_{nn} + 2d \, \text{Re}(O_{mn})  - \frac{\operatorname{tr}(O)}{d}.
\] 

Here we use $O_{mn}=O^{*}_{nm}$, and $O_{mn} = \text{Re}(O_{mn}) + i \text{Im}(O_{mn})$. 

\textbf{(3) Imaginary superposition state:}
\[
|\psi\rangle = \frac{1}{\sqrt{2}} (|m\rangle + i |n\rangle).
\]
Then:
\[
\rho = \frac{1}{2} (|m\rangle\langle m| + |n\rangle\langle n| - i |m\rangle\langle n| + i |n\rangle\langle m|),
\]
and again
\[
\sum_k \operatorname{tr}(\rho P_k) P_k = \frac{1}{2} (P_m + P_n).
\]
So:
\begin{equation}\label{constant3}
    \mathcal{M}^{-1}(\rho) = 2d \left[ \rho - \frac{d-1}{2d} (P_m + P_n) \right] - \frac{1}{d} I=P_m+P_n+d(i |n\rangle\langle m| -i |m\rangle\langle n|)- \frac{1}{d} I.
\end{equation}

Similarly, we have 
\[
\operatorname{tr}[\mathcal{M}^{-1}(\rho) \cdot O] = O_{mm} + O_{nn} - 2d \, \text{Im}(O_{mn}) - \frac{\operatorname{tr}(O)}{d}.
\]

\textbf{Conclusion:} Each of the above formulas involves only a constant number of entries from \(O = [O_{mn}]\), and requires a constant number of arithmetic operations. Hence, the classical computational cost of each post-processing step is independent of \(d\), i.e., \(O(1)\). Specifically, the calculation of \(\frac{\operatorname{tr}(O)}{d}\) can be incorporated into the final averaging step, so each post-processing requires at most four arithmetic operations. \qed

\begin{remark}[Storage complexity]
The sparse structure of $\mathcal{M}^{-1}(\rho)$ in DDB-ST also leads to efficient storage. Each measurement result can be stored as a sparse matrix with at most 4 non-trivial elements plus the constant diagonal term $-\frac{1}{d}I$. 
Specifically, we need to store:
\begin{itemize}
    \item \textbf{Case identifier}: 2 bits to distinguish between:
    \begin{itemize}
        \item Case 1: $\mathcal{M}^{-1}(\rho) = 2|t\rangle\langle t| - \frac{1}{d}I$ 
        \item Case 2: $\mathcal{M}^{-1}(\rho) = P_m + P_n + d(|m\rangle\langle n| + |n\rangle\langle m|) - \frac{1}{d}I$
        \item Case 3: $\mathcal{M}^{-1}(\rho) = P_m + P_n + d(i|n\rangle\langle m| - i|m\rangle\langle n|) - \frac{1}{d}I$
    \end{itemize}
    \item \textbf{Position indices}: at most 2 indices $(m,n)$ or $(t)$, each requiring $\lceil \log_2 d \rceil$ bits
    \item \textbf{Coefficient values}: at most 4 floating-point numbers, e.g., coefficient 2 for Case 1, $(1,1,d,d)$ for Case 2, or $(1,1,\pm id)$ for Case 3  
    \item \textbf{Universal diagonal term}: $-\frac{1}{d}$ (stored once and applied to all diagonal elements)
\end{itemize}
This results in $O(\log d)$ bits per measurement. For $m$ total measurements in DDB-ST, the total storage complexity is $O(m \log d)$ bits. 
\end{remark} 

\begin{remark}[Representation of observables]
The above constant-time result is stated with respect to the computational-basis representation, 
where both $\tilde{\rho}$ and $O$ are specified by their matrix elements. 
If instead $O$ is given in terms of its Pauli decomposition,
\[
O = \sum_{Q \in \mathcal{P}_n} \alpha_Q Q, 
\qquad \alpha_Q = \tfrac{1}{2^n}\operatorname{tr}(QO),
\]
then both Clifford-ST and DDB-ST can still evaluate each individual Pauli term $\operatorname{tr}(Q\tilde{\rho})$ in polynomial time 
(using, e.g., the Gottesman–Knill theorem or stabilizer overlap formulas). 
However, when $O$ contains exponentially many nonzero Pauli terms, 
the total post-processing cost necessarily becomes exponential due to the output size. 
Thus the constant-time advantage of DDB-ST is specific to the computational-basis representation. 
\end{remark}

Clifford-ST and DDB-ST are particularly suited for predicting global properties when $\tr(O^2)$ is bounded, e.g.\ fidelity estimation with $O=|\phi\rangle\langle \phi|$ where $\tr(O^2)=1$. 
By contrast, for a single $n$-qubit Pauli operator $Q$ one has $\tr(Q^2)=2^n$, so the variance is large unless coefficients are sufficiently small to keep $\tr(O^2)$ bounded. 

In this setting, the Pauli-ST \cite{huang2020predicting} (classical shadow tomography with random Pauli measurements) provides an efficient alternative: it can  predict $k$-local Pauli observables, where ``$k$-local" means the operator acts nontrivially on at most $k$ qubits, yielding a variance scaling as $3^k$ and making the method powerful when $k$ is small. 
More recently, the ``triply efficient shadow tomography” protocol \cite{king2025triply} shows that using two-copy joint measurements one can compress an $n$-qubit state into a $\mathrm{poly}(n)$-size classical representation from which the expectation of any chosen Pauli operator (from the full set of $4^n$ Paulis) can be extracted in $\mathrm{poly}(n)$ time, i.e., without the $k$-local restriction. Moreover, it is proved that any single-copy protocols cannot achieve sample-efficient tomography for the full Pauli set. But with any method, if one seeks to estimate exponentially many Pauli observables simultaneously, the total runtime becomes exponential due to the output size, even though each queried expectation can be obtained efficiently.

Thus, an interesting question is whether, when a general observable is specified in terms of a Pauli decomposition involving exponentially many terms,
one can design shadow-based methods whose per-sample post-processing cost remains efficient. 

\section{Proof of theoreom 2: performance guarantee}

The predicted observable is $O$. 
Denote its traceless part as $O_0=O-\mbox{tr}(O)I/d$. 
If we uniformly sample the DDBs for state $\sigma$ as introduced above, the sampling complexity is linearly dependent on the variance 
\begin{equation}
    \mathbb{E}_{U \sim \mathcal{U}} \sum_{b \in \{0, 1\}^n}   \langle b |U  \sigma U^\dag  | b\rangle \cdot \langle   b| U   \mathcal{M} ^{-1}(O_0) U^\dag | b\rangle ^2.
\end{equation}

We have  
\begin{equation}
    \mathcal{M}^{-1}(O_0)=2d\left[ O_0-\frac{d-1}{d}\sum_{k=0}^{d-1}\text{tr}(O_0P_k)P_k  \right] .
\end{equation}

Denote $\mathcal{M}^{-1}(O_0)=2d \times o$, where $o=O_0-\frac{d-1}{d}\sum_{k=0}^{d-1}\text{tr}(O_0 P_{k})P_{k}$. 

By the definition of $o$, we can calculate the relationship between its matrix elements and those of the matrix \(O_0\): 
\begin{equation} \label{eq:oO0}
\left\{\begin{aligned}
    & \text{tr}(oP_{k})=\frac{1}{d} \text{tr}(O_0P_{k})\\
   & \text{tr}(o|j\rangle\langle k|)=\text{tr}(O_0|j\rangle\langle k|).\\
\end{aligned}   \right. 
\end{equation}
This means that the diagonal elements of the operator $o$ correspond to the reciprocal of the diagonal elements of the operator $O_0$ with a denominator of $d$. 
The operators $o$ and $O_0$ share the same non-diagonal elements.   
The variance for unknown state $\sigma$ is then expressed as follows:  
 \begin{equation}\label{eq:varianceall}
     \begin{aligned} 
  \lVert O_0 \rVert_{\sigma}^2 &   = \sum_{k=0}^{d-1} \frac{2\text{tr}(\sigma P_{k}) }{2d}  \cdot \text{tr}^2(  \mathcal{M}^{-1}(O_0)  P_{k}) +\sum_{0\le j<k\le d-1} [\frac{\text{tr}(\sigma P^{+}_{jk}) }{2d}  \cdot \text{tr}^2(\mathcal{M}^{-1}(O_0)P^+_{jk}) +  \frac{\text{tr}(\sigma Q^+_{jk}) }{2d}  \cdot \text{tr}^2(\mathcal{M}^{-1}(O_0)Q^+_{jk}) ] \\
   &   +\frac{\text{tr}(\sigma P^{-}_{jk}) }{2d}  \cdot \text{tr}^2(\mathcal{M}^{-1}(O_0)P^{-}_{jk}) +  \frac{\text{tr}(\sigma Q^{-}_{jk}) }{2d}  \cdot \text{tr}^2(\mathcal{M}^{-1}(O_0)Q^{-}_{jk}) ]  \\
    &   = 4d \sum_{k=0}^{d-1} \text{tr}(\sigma P_{k}) \cdot \text{tr}^2(oP_{k})
    + 2d  \sum_{0\le j<k\le d-1} \left[  \text{tr}(\sigma P_{jk}^{\pm}) \cdot \text{tr}^2(oP_{jk}^{\pm})+\text{tr}(\sigma Q_{jk}^{\pm}) \cdot \text{tr}^2(oQ_{jk}^{\pm}) \right].
\end{aligned}
 \end{equation}

\subsection{Upper bound for the worst case}

Since $\sigma$, $P_k$, $P_{jk}^{\pm}$, and $Q_{jk}^{\pm}$ are all quantum states, their inner products satisfy $0 \le \operatorname{tr}(\sigma P_k), \operatorname{tr}(\sigma P_{jk}^{\pm}), \operatorname{tr}(\sigma Q_{jk}^{\pm}) \le 1$. 
By upper bounding each of these terms by 1, we obtain the following bound on the variance: 
  \begin{equation}\label{variance-1}
      \lVert O_0 \rVert_{\sigma}^2 \le 4d \sum_{k=0}^{d-1} \text{tr}^2(oP_{k})+2d\sum_{0\le j<k\le d-1} \left[   \text{tr}^2(oP_{jk}^{\pm})+  \text{tr}^2(oQ_{jk}^{\pm}) \right]).
  \end{equation}

Denote the matrix elements of $o$ as $o_{jk}$, where $j,k\in \{0,\cdots,d-1\}$. We rewrite Eq. (\ref{variance-1}) as $ \lVert O_0 \rVert_{\sigma}^2 \le 2d\times T$, where 
 \begin{align*}
  T   \doteq &\sum_{k=0}^{d-1} 2\text{tr}^2(oP_{k})+\sum_{0\le j<k\le d-1} \left[   \text{tr}^2(oP_{jk}^{\pm})+  \text{tr}^2(oQ_{jk}^{\pm}) \right]\\
   =& \sum_{k=0}^{d-1} 2o_{kk}^2+ \frac{1}{4}\sum_{0\le j<k\le d-1} \big[(o_{jj}+o_{kk}+o_{jk}+o_{kj})^2+  (o_{jj}+o_{kk}-o_{jk}-o_{kj})^2\\ 
   &+(o_{jj}+o_{kk}+io_{jk}-io_{kj})^2+ (o_{jj}+o_{kk}-io_{jk}+io_{kj})^2 \big] \\
   = & \sum_{k=0}^{d-1} 2o_{kk}^2+  \sum_{0\le j<k\le d-1} \big[(o_{jj}+o_{kk})^2+  2o_{jk}o_{kj} \big] \\
   \le &   \sum_{k=0}^{d-1}2 o_{kk}^2 +\sum_{0\le j<k\le d-1} \big[2(o_{jj}^2+o_{kk}^2)+  2o_{jk}o_{kj} \big]  \\
   =& 2d\sum_{k=0}^{d-1} o_{kk}^2 + \sum_{0\le j<k\le d-1}   2o_{jk}o_{kj} \\
   =& \frac{2d }{d^2}\sum_{k=0}^{d-1} [d\times o_{kk}]^2 + \sum_{0\le j<k\le d-1}   2o_{jk}o_{kj}. 
\end{align*}

When $d\ge 2$,  we have $\frac{2d }{d^2} \le 1$. So, we can deduce the upper bound of $T$. 
 \begin{align*}
  T\le \sum_{k=0}^{d-1} [d \times o_{kk}]^2 + \sum_{0\le j<k\le d-1}   2o_{jk}o_{kj} =\text{tr}(O_0^2).
\end{align*}
Here we use the relation in Eq. (\ref{eq:oO0}).

Thus the upper bound of $ \lVert O_0 \rVert_{\sigma}^2$ can be deduced. For each unknown state $\sigma$, the expectation values of $\text{tr}(\sigma P_k)$, $\text{tr}(\sigma P_{jk}^{\pm})$, and $\text{tr}(\sigma Q_{jk}^{\pm})$ are no bigger than 1. Then we have 
\begin{equation}
    \lVert O_0 \rVert_\text{ shadow}^2 = \max_{\sigma: \text{ state }}  \lVert O_0 \rVert_{\sigma }^2 \le 2dT \le  2d\times \text{tr}(O_0^2).
\end{equation}

 Consider the variance in Eq. (\ref{eq:varianceall}), the first part $4d \sum_{k=0}^{d-1} \text{tr}(\sigma P_{k}) \cdot \text{tr}^2(oP_{k})$ tends to zero as \(d\) increases. 
 As we have $ \text{tr}^2(oP_{k})=\frac{\tr^2(O_0|k\rangle\langle k|)}{d^2}\le \tr(O^2_0)/d^2$ for all $k$. Thus 
 \begin{equation}
     4d \sum_{k=0}^{d-1} \text{tr}(\sigma P_{k}) \cdot \text{tr}^2(oP_{k})\le \frac{4 \tr(O^2_0)}{d}.
 \end{equation} 

 Then for BN observable, the efficiency of sampling complexity is just related to the following  part: 
 \begin{equation}\label{equ:dia}
   V_{\mbox{diag}} = 2d \times \sum_{0\le j<k\le d-1} \left[  \text{tr}(\sigma P_{jk}^{\pm}) \cdot \text{tr}^2(oP_{jk}^{\pm})+\text{tr}(\sigma Q_{jk}^{\pm}) \cdot \text{tr}^2(oQ_{jk}^{\pm})\right]. 
 \end{equation}

 One worst case could happen when the unknown state and the observable are the same as one of the nontrivial DDB states.  
 For example, $\sigma=P_{01}^+$ and $O=P_{01}^+$. 
 Then $O_0=P_{01}^+ -I/d$, $o=\frac{d-2}{2d^2}(|0\rangle\langle 0|+|1\rangle\langle 1|)+\frac{|0\rangle\langle 1|+|1\rangle \langle 0|}{2}$. 
Thus $\tr(o P_{jk})>1/2$ and $V_{\mbox{diag}}>d$. Then in this case, the sampling complexity is linear dependent with $d$.

\subsection{Average performance analysis} 
If we sample $|\phi\rangle=U|0\rangle$ with Haar measure, the average state will be
\begin{equation}
    \int_{U(2^n)}  U|0\rangle\langle 0|U^{\dag} \mathrm{d} \mu (U)=I/2^n \;.
\end{equation}
Thus the variance of state $\sigma=I/d$ exhibits the average performance when the output $\rho$ is randomly and uniformly generated. 

 When the unknown state is $\sigma=I/d$, we have 
 \begin{equation}\label{eq:mixed}
     \tr(\sigma P_k)=\tr(\sigma P^{\pm}_{jk})=\tr(\sigma Q^{\pm}_{jk})=1/d. 
 \end{equation}
Take Eq. (\ref{eq:mixed}) into Eq. (\ref{eq:varianceall}), we have   
\begin{equation}
    \lVert O_0 \rVert_{I/d}^2 \le \frac{1 }{d} 2d T \le 2\text{tr}(O_0^2).
\end{equation} 

Thus, the average performance is efficient for bounded-norm observables. 
This value is approximately twice that of the average performance obtained by uniformly sampling from a complete set of mutually unbiased bases (MUBs), which yields $(1 + 1/2^n)\operatorname{tr}(O_0^2)$ for $d = 2^n$. 
However, a complete set of $d + 1$ MUBs is known to exist only in prime power dimensions.

\section{Proof of lemma 1: approximate average state case} 
 
If the state \(\rho\) is approximately DDB-average, then its deviation from the completely mixed state \(I/d\) is small. Specifically, we have
\begin{equation}\label{eq:approximate}
   \left|\text{tr}(\rho |\phi\rangle \langle \phi|) - \frac{1}{d}\right| \le \frac{O(\text{poly}(\log d))}{d} 
\end{equation}
for all snapshots \(|\phi\rangle \in \mathcal{S}_{\text{DDB}}\). This implies that \(\text{tr}(\rho |\phi\rangle \langle \phi|) \le \frac{O(\text{poly}(\log d)) + 1}{d}\). 
Substituting this into Eq. (\ref{eq:varianceall}), we can deduce
\begin{equation}\label{eq:varianceaverage}
\begin{aligned} 
\lVert O_0 \rVert_{\rho}^2 &= 4d \sum_{k=0}^{d-1} \text{tr}(\rho P_{k}) \cdot \text{tr}^2(o P_{k})
+ 2d \sum_{0 \le j < k \le d-1} \left[ \text{tr}(\rho P_{jk}^{\pm}) \cdot \text{tr}^2(o P_{jk}^{\pm}) + \text{tr}(\rho Q_{jk}^{\pm}) \cdot \text{tr}^2(o Q_{jk}^{\pm}) \right] \\
&\le O(\text{poly}(\log d) + 1) \times 2T \\
&\le O(\text{poly}(\log d)) \times \text{tr}(O^2_0).
\end{aligned}
\end{equation}

Denote \(\epsilon_1 = \frac{O(\text{poly}(\log d))}{d}\). It is an interesting question to characterize the proportion of randomly chosen states for various levels of deviation \(\epsilon_1\), such as \(\frac{\log d}{d}\), \(\frac{\log^2 d}{d}\), \(\frac{\log^3 d}{d}\), and so forth.

\subsection{More exact variance formula}
 
We now present a more concise expression for \(V_{\text{diag}}\) in Eq. (\ref{equ:dia}) by substituting the form given in Eq. (\ref{equ:four}). The variance for \(\rho\) and a BN-observable \(O\), given by \(\lVert O_0 \rVert_{\sigma}^2\), can be expressed as \(V_{\text{diag}}\) plus a term that vanishes as \(d \to \infty\). Hence, the growth of \(\lVert O_0 \rVert_{\sigma}^2\) is determined by the behavior of \(V_{\text{diag}}\). If \(V_{\text{diag}}\) scales polynomially, then \(\lVert O_0 \rVert_{\sigma}^2\) also exhibits polynomial scaling; conversely, if \(V_{\text{diag}}\) scales exponentially, \(\lVert O_0 \rVert_{\sigma}^2\) will follow an exponential growth pattern as well.

\begin{equation}\label{equ:diadetailed}
\begin{aligned}
    V_{\text{diag}} &= 2d \sum_{0\le j<k\le d-1} \left[ \operatorname{tr}(\sigma P_{jk}^{\pm}) \cdot \operatorname{tr}^2(o P_{jk}^{\pm}) + \operatorname{tr}(\sigma Q_{jk}^{\pm}) \cdot \operatorname{tr}^2(o Q_{jk}^{\pm}) \right] \\
    &= \frac{d}{4} \sum_{0\le j<k\le d-1} \Big[ (\sigma_{jj} + \sigma_{kk} + \sigma_{jk} + \sigma_{kj})(o_{jj} + o_{kk} + o_{jk} + o_{kj})^2 \\
    &\quad + (\sigma_{jj} + \sigma_{kk} - \sigma_{jk} - \sigma_{kj})(o_{jj} + o_{kk} - o_{jk} - o_{kj})^2 \\
    &\quad + (\sigma_{jj} + \sigma_{kk} + i\sigma_{jk} - i\sigma_{kj})(o_{jj} + o_{kk} + i o_{jk} - i o_{kj})^2 \\
    &\quad + (\sigma_{jj} + \sigma_{kk} - i\sigma_{jk} + i\sigma_{kj})(o_{jj} + o_{kk} - i o_{jk} + i o_{kj})^2 \Big] \\
    &= \frac{d}{4} \sum_{0\le j<k\le d-1} \Big[ 4(\sigma_{jj} + \sigma_{kk})(o_{jj} + o_{kk})^2 \\
    &\quad + 4(\sigma_{jk} + \sigma_{kj})(o_{jj} + o_{kk})(o_{jk} + o_{kj}) \\
    &\quad - 4(\sigma_{jk} - \sigma_{kj})(o_{jj} + o_{kk})(o_{jk} - o_{kj}) \\
    &\quad + 8(\sigma_{jj} + \sigma_{kk}) o_{jk} o_{kj} \Big]\\
        &= d \sum_{0\le j<k\le d-1} [ (\sigma_{jj} + \sigma_{kk})(o_{jj} + o_{kk})^2 + (\sigma_{jk} + \sigma_{kj})(o_{jj} + o_{kk})(o_{jk} + o_{kj}) - (\sigma_{jk} - \sigma_{kj})(o_{jj} + o_{kk})(o_{jk} - o_{kj}) \\
    &\quad+ 2(\sigma_{jj} + \sigma_{kk}) o_{jk} o_{kj} ].
\end{aligned}
\end{equation}

Now, consider the case where \(O\) is an off-diagonal observable, meaning \(O_{jj} = o_{jj} = 0\) for all \(j = 0, \ldots, d-1\). Under this condition, the expression simplifies to:
\[
V_{\text{diag}} = 2d \sum_{0 \le j < k \le d-1} (\sigma_{jj} + \sigma_{kk}) o_{jk} o_{kj}.
\]

\section{Efficient estimation of stabilizer states with DDB-ST}

Stabilizer states are not only mathematically structured but also physically central, as they form the basis of stabilizer quantum error correcting codes (including CSS codes, surface codes, toric codes, and quantum LDPC codes), play a key role in fault-tolerant quantum computation, and serve as free states in magic state resource theory. 

Property 2 in the main text is easily verified through direct calculation.

\begin{lemma} 
    Given arbitrary \( n \)-qubit stabilizer state \( |\Psi\rangle \)  in Eq. (\ref{eq:stabilizer}), we can efficiently construct a Clifford circuit \( T \), composed of elementary gates from the set \(\{S, \text{CZ}, \text{CX}\}\), such that \( |\Phi\rangle = T |\Psi\rangle = |0\rangle^{\otimes (n-r)} \otimes |\Phi_r\rangle \). Here, \(\text{CZ}\) and \(\text{CX}\) denote the controlled-Z and controlled-X gates, respectively, and \( |\Phi_r\rangle \) is an \( r \)-qubit stabilizer state with \( 2^r \) nonzero amplitudes.
\end{lemma}

\textit{Proof:} By Eq. (42) of \cite{bravyi2019simulation}, any stabilizer state can be expressed as
\[
|\Psi\rangle = \omega U_C U_H |s\rangle,
\]
where \( U_C \) and \( U_H \) are \( C \)-type and \( H \)-type Clifford operators, \( s \in \{0, 1\}^n \) is a basis vector, and \( \omega \) is a complex phase factor. The \( C \)-type Clifford operators are those that can be decomposed into gates from the set \(\{S, \text{CZ}, \text{CX}\}\), while the \( H \)-type Clifford operators only consist of Hadamard gates. Furthermore, a polynomial-time algorithm is provided to efficiently obtain this CH-form of the stabilizer state using the stabilizer tableaux representation.

Consequently, the target Clifford circuit \( T \) can be decomposed into \( U_C^{\dagger} \), followed by a sequence of SWAP operations (each equivalent to three \(\text{CX}\) gates). \qed

\subsection{Proof of Theorem 3}

We prove Theorem 3: for any $n$-qubit stabilizer state $|\Psi\rangle$ and BN-observable $O$, the expectation value $\tr(|\Psi\rangle\langle\Psi| O)$ can be efficiently estimated using DDB-ST with $O(\text{poly}(n))$ samples and computational resources.

\begin{proof} 
We categorize the cases based on the number of non-zero coefficients present in \( |\Psi\rangle \langle \Psi| \) by Eq. (\ref{eq:stabilizer}).


\textbf{Case 1:} \( n - \log(\operatorname{poly}(n)) \leq r \leq n \).  
The stabilizer state \( |\Psi\rangle \) is approximately DDB-average. Thus, \( O(\operatorname{poly}(n)) \) samples and computational resources are sufficient for efficient estimation using \( n \)-qubit DDB-ST (Lemma~1).

\textbf{Case 2:} \( 0 \le r \le \log(\operatorname{poly}(n)) \).  
The matrix \( |\Psi\rangle \langle \Psi| \) is sparse, containing at most \( 4^r \) non-zero coefficients, making direct computation of \( \mathrm{tr}(|\Psi\rangle \langle \Psi| O) \) efficient.

\medskip
Now we consider the general $r$. 

\textbf{Proof strategy overview:} 
The key insight is to exploit the structure of stabilizer states to reduce 
the estimation problem to a smaller dimensional space. Our approach proceeds 
in three main steps: (1) \textbf{Dimensional reduction}: Transform the 
$n$-qubit stabilizer state to concentrate all quantum coherence in only $r$ 
qubits, where $r$ is the stabilizer rank. (2) \textbf{Efficient estimation}: 
Apply $r$-qubit DDB-ST on the reduced problem, leveraging that $r$-qubit 
stabilizer states are approximately DDB-average. (3) \textbf{Error control}: 
The total estimation error is $\epsilon + \sqrt{\operatorname{tr}(O^2)}/\sqrt{2^r}$, 
where $\epsilon$ is the DDB-ST sampling error and the second term arises from 
neglecting certain diagonal contributions in the post-processing.

\textbf{State transformation.} 
For a given stabilizer state $|\Psi\rangle$, we construct its representation in two main steps to enable efficient DDB-ST estimation.

The construction proceeds as follows:
\begin{enumerate}
    \item \textbf{Stabilizer decomposition}: Using stabilizer tableau algorithms, find a Clifford circuit $V$ and computational basis state $|j\rangle$ such that $|\Psi\rangle = V|j\rangle$, where $V$ decomposes into standard Clifford generators (Hadamard, Phase, and CNOT gates), requiring $O(n^3)$ time with at most $O(n^2)$ gates.
    
    \item \textbf{CH-form conversion}: Transform the circuit $V$ into CH-form~\cite{bravyi2019simulation}, yielding $|\Psi\rangle = \omega U_C U_H |s\rangle$, where $U_C$ contains only $\{S, \text{CZ}, \text{CX}\}$ gates and $U_H$ contains only Hadamard gates. This conversion requires runtime $O(n)$ per $S$, CZ, and CX gate, and $O(n^2)$ per Hadamard gate.
\end{enumerate}

The total time complexity for obtaining the CH-form is at most $O(n^4)$
(a conservative worst-case bound for each gate), and can be reduced in practice with
implementation optimizations.

By Lemma~S1, we can further construct a unitary operator $T$ (composed of $\{S,CZ,CX\}$ gates) such that
\[
|\Phi\rangle = T|\Psi\rangle
\]
transforms the state to the desired form:
\begin{equation}\label{eq:permutation}
\begin{aligned}
|\Phi\rangle & = |0\rangle^{\otimes (n-r)} \otimes \frac{1}{\sqrt{2^r}} \sum_{k \in \{0,1\}^r} \omega^{q'(k)} |k\rangle \\
&= |0\rangle^{\otimes (n-r)} \otimes |\Phi_r\rangle.
\end{aligned}
\end{equation}

The transformation $T$ effectively rearranges qubits so that all quantum coherence is concentrated in the last $r$ qubits, while the first $n-r$ qubits are in the $|0\rangle$ state. 
In other words, we obtain a new stabilizer state \( |\Phi\rangle \) where only the first \( 2^r \) components are non-zero. 
This is achieved through $U_C^{\dagger}$ operations and SWAP gates as described in Lemma~S1. 
 And once the CH-form is known, $T$ is immediately obtained. 
 Importantly, this preprocessing is performed only once: in the subsequent estimation protocol we merely use the monomial structure of $T$ (permutation plus phase) to track basis states, which adds no extra overhead beyond polynomial factors.

\medskip
\textbf{Reduction to \( r \)-qubit DDB-ST.}
The target expectation value can be rewritten as
\begin{equation}
\begin{aligned}
    \mathrm{tr}(|\Psi\rangle \langle \Psi| O) 
    &= \mathrm{tr}(T|\Psi\rangle \langle \Psi| T^{\dagger} \cdot T O T^{\dagger}) \\
    &= \mathrm{tr}(|\Phi\rangle \langle \Phi| \cdot T O T^{\dagger}),
\end{aligned}
\end{equation}
where \( |\Phi\rangle = |0\rangle^{\otimes (n-r)} \otimes |\Phi_r\rangle \), and \( |\Phi_r\rangle \) is an \( r \)-qubit DDB-average stabilizer state (by Property~2).

Since the matrix form of \( |\Phi\rangle \langle \Phi| \) is block-diagonal with a single nonzero block in the upper-left corner, we have:
\[
|\Phi\rangle \langle \Phi| = 
\begin{bmatrix}
|\Phi_r\rangle \langle \Phi_r| & \mathbf{0} \\
\mathbf{0} & \mathbf{0}
\end{bmatrix},
\]
and hence only the leading \( 2^r \times 2^r \) submatrix of \( T O T^{\dagger} \), denoted by \([T O T^{\dagger}]_{2^r \times 2^r}\), contributes to the trace. Therefore,
\[
\mathrm{tr}(|\Psi\rangle \langle \Psi| O) = \mathrm{tr}(|\Phi_r\rangle \langle \Phi_r| \cdot [T O T^{\dagger}]_{2^r \times 2^r}).
\]
Since unitary conjugation preserves the Hilbert-Schmidt norm, we have \( \operatorname{tr}[(T O T^{\dagger})^2] = \operatorname{tr}(O^2) \), which implies
\[
\operatorname{tr}([T O T^{\dagger}]_{2^r \times 2^r}^2) \le \operatorname{tr}(O^2).
\]

We can apply \( r \)-qubit DDB-ST to estimate this quantity efficiently with variance bounded by \( O(1) \cdot \operatorname{tr}(O^2) \),  where the constant factor arises from the reduction of the numerator in Eq.~\eqref{eq:approximate} from \( O(\mathrm{poly}(\log d)) \) to 1.

\medskip
\textbf{Post-processing in $r$-qubit DDB-ST.}
In the post-processing step, we draw 
\[
S = O\!\left(\frac{\operatorname{tr}(O^2)}{\epsilon^2}\log \frac{1}{\sigma}\right)
\]
independent samples and for each sample we evaluate a value of the form
\(
\mathrm{tr}(\mathcal{M}^{-1}(|\phi_r\rangle\langle\phi_r|)\,[T O T^{\dagger}]_{2^r\times 2^r})
\).
where each \( |\phi_r\rangle \) is an \( r \)-qubit DDB state, and \( \mathcal{M}^{-1} \) is the inverse channel defined in Eq.~\eqref{inverse}, with \( d = 2^r \). Let \( \tau_r := \mathcal{M}^{-1}(|\phi_r\rangle \langle \phi_r|) \). By Eqs.~\eqref{constant1}, \eqref{constant2}, and \eqref{constant3}, we have
\begin{equation}
    \tau_r = \tau_r' - \frac{I_r}{2^r},
\end{equation}
where \( \tau_r' \) is a \( 2^r \)-dimensional operator with at most four nonzero components, and \( I_r \) is the identity on \( \mathbb{C}^{2^r} \).

Although \([T O T^{\dagger}]_{2^r \times 2^r}\) is a submatrix of \( T O T^{\dagger} \), its explicit form is not required. Define
\[
\tau := (|0\rangle \langle 0|)^{\otimes (n - r)} \otimes \tau_r,
\]
then we obtain
\begin{equation}\label{eq:transform}
\begin{aligned}
\mathrm{tr}(\tau_r \cdot [T O T^{\dagger}]_{2^r \times 2^r}) 
&= \mathrm{tr}(\tau \cdot T O T^{\dagger}) 
= \mathrm{tr}(T^{\dagger} \tau T \cdot O) \\
&= \mathrm{tr}\big(T^{\dagger} \big[(|0\rangle \langle 0|)^{\otimes (n - r)} \otimes \tau_r\big] T \cdot O\big) \\
&= \mathrm{tr}\big(T^{\dagger} \big[(|0\rangle \langle 0|)^{\otimes (n - r)} \otimes \tau_r'\big] T \cdot O\big)
- \frac{1}{2^r} \cdot \mathrm{tr}\big(T^{\dagger} \big[(|0\rangle \langle 0|)^{\otimes (n - r)} \otimes I_r\big] T \cdot O\big).
\end{aligned}
\end{equation}

The first term in the subtraction of Eq.~\eqref{eq:transform} is defined as 
\begin{equation} \label{eq:l1}
    L_1:=\mathrm{tr}\big(T^{\dagger} \big[(|0\rangle \langle 0|)^{\otimes (n - r)} \otimes \tau_r'\big] T \cdot O\big). 
\end{equation}

The right-hand side of the subtraction in Eq.~\eqref{eq:transform} is defined as 
\begin{equation}\label{eq:l2}
    L_2 := \frac{1}{2^r} \, \mathrm{tr}\left(T^{\dagger} \left[(|0\rangle\langle 0|)^{\otimes (n-r)} \otimes I_r\right] T \cdot O\right). 
\end{equation} 
In post-processing, we only calculate the value of $L_1$ in Eq. (\ref{eq:l1}) and neglect the value of $L_2$ in Eq. (\ref{eq:l2}). 

\medskip
\textbf{Time complexity to calculate $L_1$ in Eq.~(\ref{eq:l1}).} 
Since $\tau_r'$ contains at most four nonzero components, the tensor product 
$(|0\rangle \langle 0|)^{\otimes (n-r)} \otimes \tau_r'$ is also extremely sparse, 
with at most four nonzero entries. 
Recall that $T$ is a Clifford circuit composed of gates from $\{S, CZ, CX\}$, 
and hence it is a \emph{monomial Clifford operator}: on computational basis states 
it acts as a permutation together with a phase factor. 
Thus conjugating a rank-one operator $|n\rangle\langle m|$ by $T$ yields 
\[
T|n\rangle\langle m|T^\dagger 
= e^{i(\phi(n)-\phi(m))}\,|\pi(n)\rangle\langle \pi(m)|,
\]
where $\pi(\cdot)$ is a permutation of basis strings and $\phi(\cdot)$ is a quadratic phase function. 
Both $\pi(x)$ and $\phi(x)$ can be evaluated in $O(n^2)$ time in the worst case 
(since $\pi(x)$ is an affine linear transformation and $\phi(x)$ is a quadratic form over $\mathbb{F}_2$). 
Consequently, 
\[
T^{\dagger}\!\left[(|0\rangle\langle 0|)^{\otimes (n-r)}\otimes \tau_r'\right]T
\]
is a sum of at most four such rank-one terms, obtained by at most four evaluations of $T|n\rangle$ or $(T|m\rangle)^\dagger$.  

Therefore, once \( T^{\dagger} \left[(|0\rangle \langle 0|)^{\otimes (n - r)} \otimes \tau_r'\right] T \) is obtained, the final trace 
\[
\mathrm{tr}\left(T^{\dagger} \left[(|0\rangle \langle 0|)^{\otimes (n - r)} \otimes \tau_r'\right] T \cdot O\right)
\]
can be evaluated in constant time, due to the sparsity of the operator involved. Hence the overall classical cost for a single measurement to compute $L_1$ is $O(n^2)\cdot O(1)$.

\textbf{Overall complexity.}
The total runtime decomposes as 
\begin{equation}
T_{\mathrm{total}} = T_{\mathrm{prep}} \;+\; S \cdot (T_{\text{prep-state}} + T_T + T_{\mathrm{classical}}),
\end{equation}
where
\begin{itemize}
    \item $T_{\mathrm{prep}} = O(n^4)$ 
is the one-time preprocessing cost to obtain the CH-form and the Clifford circuit $T$. 
This preprocessing is carried out once in advance and does not repeat for each measurement. 
In practice, more efficient implementations may reduce this bound, 
but $O(n^4)$ suffices as a conservative estimate. 

    \item $S = O(1) \cdot \mathrm{tr}(O^2) \cdot \frac{\log(1/\sigma)}{\epsilon^2}$ 
    is the number of samples required by $r$-qubit DDB-ST;
    \item $T_{\text{prep-state}}$ is the physical cost of preparing the stabilizer state $|\Psi\rangle$ once per sample;
  \item $T_T = O(n^2)$ is the per-sample cost of physically applying 
the Clifford circuit $T$ to the stabilizer state $|\Psi\rangle$ 
in order to obtain the transformed state $|\Phi_r\rangle$, 
on which the $r$-qubit DDB-ST measurement is performed.  
    \item $T_{\mathrm{classical}} = O(n^2)$ is the calculation cost in each $r$-qubit DDB-ST measurements.  As discussed above, it is mainly the cost of evaluating the affine permutation $\pi(\cdot)$ and quadratic phase $\phi(\cdot)$ 
    for each measurement outcome.
\end{itemize}

Thus,
\begin{equation}\label{eq:complexity}
 T_{\mathrm{total}} = O(n^4) 
+ O\!\left(\frac{\tr(O^2)}{\epsilon^2}\log \tfrac{1}{\sigma}\right)\cdot 
\big(T_{\text{prep-state}} + O(n^2)\big).
    \end{equation}


\medskip
\textbf{Error by neglecting $L_2$ in Eq. (\ref{eq:l2}).}
The value of $L_2$ in Eq. (\ref{eq:l2}) can be simplified to \(\frac{1}{2^r} \sum_{i \in A} O_{ii} \), where \( A \subset \{0,1,\dots,2^n - 1\} \) is an index set of size \( |A| = 2^r \) corresponding to the diagonal positions labeled by the support of 
\[
T^{\dagger} \left[(|0\rangle\langle 0|)^{\otimes (n-r)} \otimes I_r \right] T.
\]

Note that \( (|0\rangle\langle 0|)^{\otimes (n-r)} \otimes I_r \) is a diagonal projector with exactly \( 2^r \) ones on the diagonal and zeros elsewhere. Since \( T \) is composed of gates in  $\{S,CZ,CX\}$, we have $
T^{\dagger} \left[(|0\rangle\langle 0|)^{\otimes (n-r)} \otimes I_r \right] T
$ is then also a diagonal operation with $2^r$ ones determined by $T$ and zeros at other places. 

Thus $L_2$ in Eq. (\ref{eq:l2}) is equal to \( \frac{1}{2^r} \) times the sum of \( 2^r \) diagonal elements of the \( 2^n \times 2^n \) matrix \( O \), where the selected positions are determined by the support of the permuted projector \( T^{\dagger} \left[(|0\rangle\langle 0|)^{\otimes (n-r)} \otimes I_r\right] T \).

In the general case, the second term can be bounded by
\begin{equation}\label{eq:l2bound}
    |L_2|\;\le\;\frac{1}{2^r}\,\bigg|\sum_{i\in A} O_{ii}\bigg|
\;\le\;\frac{\sqrt{\mathrm{tr}(O^2)}}{\sqrt{2^r}}.
\end{equation}
This bound can be shown as follows. Without loss of generality, assume that the nonzero positions of the diagonal projector 
\(
T^{\dagger} \left[(|0\rangle\langle 0|)^{\otimes(n-r)} \otimes I_r \right] T
\)
correspond to the first \( 2^r \) diagonal entries of \( O \). Then we can write
\[
L_2 = \frac{1}{2^r} \sum_{i=1}^{2^r} O_{ii}.
\]

By the Cauchy--Schwarz inequality, we have
\[
|\sum_{i=1}^{2^r} O_{ii} |
\le \sqrt{2^r} \cdot \left( \sum_{i=1}^{2^r} O_{ii}^2 \right)^{1/2} 
\le \sqrt{2^r} \cdot \left( \sum_{i=1}^{2^n} O_{ii}^2 \right)^{1/2} 
\le \sqrt{2^r} \cdot \sqrt{\operatorname{tr}(O^2)}.
\]

Dividing both sides by \( 2^r \), we obtain the relation in Eq. (\ref{eq:l2bound}). 

In the general case with arbitrary \( r \), the estimation error is bounded by 
\[
\epsilon + \frac{\sqrt{\operatorname{tr}(O^2)}}{\sqrt{2^r}}. 
\] with the time complexity in Eq. (\ref{eq:complexity}).

The term \( \epsilon \) accounts for the estimation error introduced by the DDB-ST procedure. The term $\frac{\sqrt{\operatorname{tr}(O^2)}}{\sqrt{2^r}}$ arises from neglecting the contribution of \( L_2 \) in each estimation step.

\medskip
\textbf{Impact of the error term \( \frac{\sqrt{\operatorname{tr}(O^2)}}{\sqrt{2^r}} \).}
Consider the special case where \( O \) is an off-diagonal observable with vanishing diagonal elements. In this case, \( L_2 \) in Eq.~\eqref{eq:l2} evaluates to zero, since all diagonal entries \( O_{ii} \) are zero. 
Consequently, the total estimation error reduces to \( \epsilon \), and is independent of the term \( \frac{\sqrt{\operatorname{tr}(O^2)}}{\sqrt{2^r}} \). 

For the general case, we neglect \( L_2 \)  in Eq.~\eqref{eq:l2}. 
While for the standard DDB-ST applied to the full \( n \)-qubit system, the corresponding correction term simplifies to \( \operatorname{tr}(O)/2^n \), which can be computed directly since \( \operatorname{tr}(O) \) is assumed to be known.

When \( r \) is small—for example, \( 0 \le r \le \log(\mathrm{poly}(n)) \)—the term \( \frac{\sqrt{\operatorname{tr}(O^2)}}{\sqrt{2^r}}\) can have a significant impact on the overall estimation error. However, in this regime, the state \( |\Psi\rangle\langle\Psi| \) is sparse, and the exact computation of \( \operatorname{tr}(|\Psi\rangle\langle\Psi| O) \) remains tractable through direct calculation.

In contrast, when \( r \) becomes large—for instance, \( r = O(n) \)—the term \( \frac{\sqrt{\operatorname{tr}(O^2)}}{\sqrt{2^r}} \) becomes negligible as we have \( \operatorname{tr}(O^2) \le \operatorname{poly}(n) \) for BN observable $O$. Although exact computation of \( \operatorname{tr}(|\Psi\rangle\langle\Psi| O) \) is no longer feasible in this case, the error introduced by neglecting the \( L_2 \) term remains well-controlled due to the exponential suppression from the denominator \( 2^r \). And it is efficient to perform the $r$-qubit DDB-ST on the transformed state.

\end{proof}

\subsection{Practical considerations and limitations}  
It is worth noting that the above analysis  assumes the stabilizer state \( |\Psi\rangle \) is known explicitly. In this case, the most straightforward way to estimate \( \mbox{tr}(|\Psi\rangle\langle \Psi| O) \) for a BN observable \( O \) is to directly perform the physical measurement corresponding to \( O \) on \( |\Psi\rangle \), and then collect statistics from the measurement outcomes. However, in practice, the observable \( O \) may be difficult to implement physically, or we may wish to estimate the expectation values of multiple observables \(\{ O_k \}\) simultaneously.
In such cases, classical shadow tomography offers a significant advantage: a single set of shadow measurement data can be reused to estimate \( \operatorname{tr}(|\Psi\rangle\langle\Psi| O_k) \) for all \( L \) observables with total sample complexity 
\(
m = O\!\left(\max_k \|O_k\|_{\text{shadow}}^2 \cdot \frac{\log(L/\delta)}{\varepsilon^2}\right),
\)
which scales only logarithmically with \( L \). In contrast, direct measurement would require \( O(L/\varepsilon^2) \) samples total, scaling linearly with the number of observables. This data reusability becomes increasingly valuable as \( L \) grows large, allowing shadow tomography to amortize its measurement cost across multiple estimation tasks. For instance, when characterizing a quantum device or algorithm, one often needs to evaluate hundreds or thousands of different observables on the same quantum state, making the ``measure once, estimate many" paradigm of shadow tomography particularly advantageous.
 
\textbf{Infeasibility of hybrid Clifford-ST and DDB-ST approaches.}  
Finally, we would like to point out that it is not feasible to combine the techniques of Clifford-ST and DDB-ST to efficiently estimate \( \operatorname{tr}(\rho O) \) for   arbitrary \( n \)-qubit quantum state \( \rho \)  and any BN observable \( O \)  observable with known trace.
 
  The main bottleneck of Clifford-ST lies in its post-processing cost, namely, it does not guarantee that \( \operatorname{tr}(U_k^\dagger |j\rangle \langle j| U_k O) \) can always be computed in polynomial time for each single-shot measurement. While \( U_k^\dagger |j\rangle \langle j| U_k \) is a stabilizer state and \( O \) is a BN observable, which seemingly allows for efficient estimation of \( \operatorname{tr}(U_k^\dagger |j\rangle \langle j| U_k O) \) by DDB-ST as a subroutine (Theorem 3), the issue lies in the accumulated error.
 
  Specifically, in the post-processing formula of Clifford-ST (Eq.~\ref{eq:post-processing2}), each term \( \operatorname{tr}(U_k^\dagger |j\rangle \langle j| U_k O) \) is multiplied by a coefficient of \( 2^n + 1 \). As a result, the DDB-ST subroutine used to estimate \( \operatorname{tr}(U_k^\dagger |j\rangle \langle j| U_k O) \) must achieve exponentially small error in order to ensure the overall accuracy of the weighted term \( (2^n + 1) \operatorname{tr}(U_k^\dagger |j\rangle \langle j| U_k O) \). This, in turn, requires an exponential number of samples in DDB-ST, which severely limits the efficiency of the estimation and ultimately negates the potential advantages of such a hybrid approach.

\end{document}